\documentclass[%
twocolumn, prl, aps, superscriptaddress, longbibliography,showpacs,amsmath,amssymb,floatfix
]{revtex4-1}

\usepackage{graphicx}
\usepackage{dcolumn}
\usepackage{bm}
\usepackage{graphicx}                                               
\usepackage{amssymb}

\usepackage{amsmath}
\usepackage{epsfig}
\usepackage{xcolor}
\usepackage{tabu}
\usepackage{mathtools}
\usepackage[colorlinks,linkcolor=blue,anchorcolor=blue,citecolor=blue,urlcolor=blue]{hyperref}
\usepackage{physics}
\usepackage{float}
\usepackage{diagbox}
\usepackage{inputenc}

\begin{document}

\preprint{APS/123-QED}

\title{Theory of mobility edge and non-ergodic extended phase in coupled random matrices}
\author{Xiaoshui Lin}
\affiliation{CAS Key Laboratory of Quantum Information, University of Science and Technology of China, Hefei, 230026, China}

\author{Guang-Can Guo}
\affiliation{CAS Key Laboratory of Quantum Information, University of Science and Technology of China, Hefei, 230026, China}
\affiliation{Synergetic Innovation Center of Quantum Information and Quantum Physics, University of Science and Technology of China, Hefei, Anhui 230026, China}
\affiliation{Hefei National Laboratory, University of Science and Technology of China, Hefei 230088, China}

\author{Ming Gong}
\email{gongm@ustc.edu.cn}
\affiliation{CAS Key Laboratory of Quantum Information, University of Science and Technology of China, Hefei, 230026, China}
\affiliation{Synergetic Innovation Center of Quantum Information and Quantum Physics, University of Science and Technology of China, Hefei, Anhui 230026, China}
\affiliation{Hefei National Laboratory, University of Science and Technology of China, Hefei 230088, China}

\date{\today}

\begin{abstract}
The mobility edge, as a central concept in disordered models for localization-delocalization transitions, has rarely been discussed in the context of random matrix theory (RMT). 
Here we report a new class of random matrix model by direct coupling between two random matrices, showing that their overlapped spectra and un-overlapped spectra exhibit totally different scaling behaviors, which can be used to construct tunable mobility edges.
This model is a direct generalization of the Rosenzweig-Porter model, which hosts ergodic, localized, and non-ergodic extended (NEE) phases.
A generic theory for these phase transitions is presented, which applies equally well to dense, sparse, and even corrected random matrices in different ensembles. 
We show that the phase diagram is fully characterized by two scaling exponents, and they are mapped out in various conditions. 
Our model provides a general framework to realize the mobility edges and non-ergodic phases in a controllable way in RMT, which pave avenue for many intriguing applications both from the pure mathematics of RMT and the possible implementations of ME in many-body models, chiral symmetry breaking in QCD and the stability of the large ecosystems.  
\end{abstract}

\maketitle


Random matrix theory (RMT) dealing with statistics of eigenvalues and observations (associated with the eigenvectors) in large random matrices \citep{Forrester2010LogGas, Brody1981Randommatrix, Mehta2004Random, Livan2018Introduction} is one of the most interdisciplinary fields in physics and probability theory.
Initially developed by Wigner and Dyson in the analysis of nucleus spectra \citep{Wigner1955Characteristic, dyson1962brownianmotion}, the RMT has since found numerous applications in different fields, including quantum chaos \citep{Bohigas1982Characterization, Berry1977Level, Kos2018Many-body}, Anderson localization \citep{Mirlin2000Statistics, Goda2006InverseAL}, many-body localization (MBL) \citep{Pal2010Many-body, Nandkishore2015Many-body, Abanin2019Colloquium, Alet2018Many-body, Imbrie2016Diagonalization}, quantum chromodynamics (QCD) \cite{Stephanov1996RandomMatrixModel, Verbaarschot1994Spectrum, Kieburg2013Spectral, Verbaarschot2000Random} and ecological stability \cite{May1972Will, Servan2018Coexistence, Grilli2017Feasibility}, {\it etc.}. 
However, the central concept of mobility edge (ME) for energy-resolved localization-delocalization transition, which has been widely explored in the single particle disordered models \cite{Mott1987Mobility, Semeghini2015Measurement, Pixley2015Anderson, Biddle2009Localization, Biddle2010Predicted, Ganeshan2015Nearest, Wang2020one-dimensional, WangObservation2022, Alex2021Interaction}, 
and many-body disordered models \citep{Wei2019Investigating, Luitz2015Many-body, Brighi2020Stability, Deng2017Many-body, Chanda2020Many-body, Nag2017many-body, Modak2015Many-body, Deroeck2016absence, Deroeck2017Stability, Zhang2022Localization, Lazarides2015Fate, Luschen2018Single-particle, Kohlert2019Observation}, has rarely been discussed in the context of RMT.
To the best of our knowledge, it has only been discussed recently in the L\'{e}vy matrices \cite{Cizeau1994Theory, tarquini2016level, Biroli2021Levy, aggarwal2019goe, Sarkar2023Tuning}. 
However, this model, with a heavy-tail distribution, is much more challenge to be realized in the current experiments, in which the uniform and Gaussian random variables are much more appealing. 
Therefore, as an open question, the general construction of random matrices for tunable ME without a heavy-tail distribution and the underlying mechanism for ME is still unclear, which also limits our understanding of many-body ME in realistic physical models \cite{Deroeck2016absence, Deroeck2017Stability, Zhang2022Localization, Lazarides2015Fate, Luschen2018Single-particle, Kohlert2019Observation}.

\begin{figure}[hbtp!]
\centering
\includegraphics[width=0.38\textwidth]{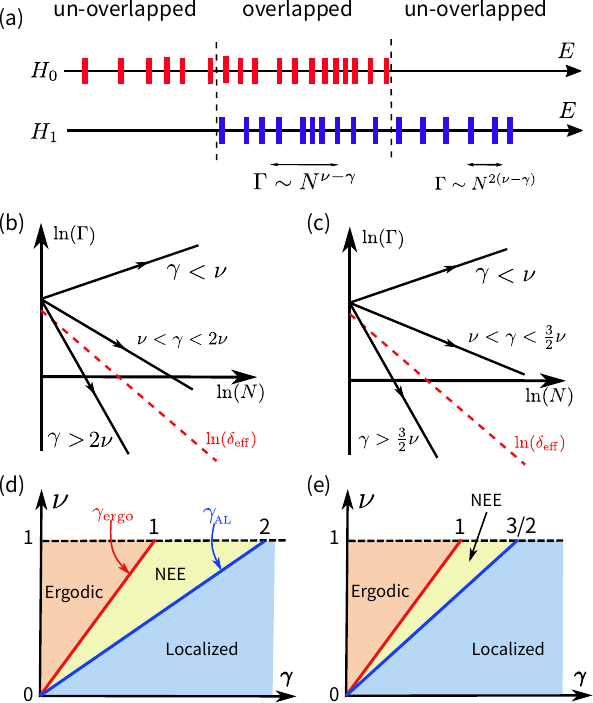}
\caption{
(a) Overlapped and un-overlapped spectra between $H_0$ and $H_1$.
The states in the energy window $[E - \Gamma/2, E + \Gamma/2] $ are hybridized by the off-diagonal block coupling.
(b)-(c) The scaling of $\Gamma$ versus system size $N$ in the overlapped (b) and un-overlapped spectra (c).
(d)-(e) The phase diagram of the model (\ref{eq-coupled-model}) for overlapped spectra (d); and un-overlapped spectra (e), with $\gamma_\text{ergo}$ ($\gamma_\text{AL}$) being the boundary between the NEE phase and ergodic (localized) phase.}
\label{fig-schematic}
\end{figure}

Here, we propose a coupled random matrix model for the realization of ME and the non-ergodic extended (NEE) phase. 
This model with a block structure is constructed from two independent random matrices with their coupling strength scaled as $N^{-\gamma/2}$ and coupling connectance scaled as $N^{\nu}$, where $\gamma$ and $\nu$ are two major parameters to characterize all phases.  
The motivation behind this construction is three-fold:
(I) With only direct coupling between different random matrices, the overlapped and un-overlapped spectra would exhibit different scaling behaviors (see Fig. \ref{fig-schematic} (a)), giving rise to ME. This is different from the RP ensemble with only one single block (see S1 and S2 in Ref. \cite{SI}).
(II) The coupling will become irrelevant when $\gamma$ is large and relevant when $\gamma$ is small (see Fig. \ref{fig-schematic} (b) - (c)), giving rise to ergodic, localized, and NEE phases by 
varying of $\gamma$ and $\nu$, similar to the Rosenzweig-Porter ensemble \cite{Rosenzweig1960REpulsion, Kravtsov2015Random, Bogomolny2018Eigenfunction, Facoetti2016Non-ergodic, Khaymovich2020Fragile, Biroli2021Levy, Tomasi2022Non-Hermitian, Wouter2022Circular} and $\beta$-ensemble \cite{Das2022Nonergodic, das2023absence}.
(III) The block structure of our model is closely related to the Hamiltonian of many-body disordered models, in which the block structure is due to the symmetry \cite{Giraud2022Probing}. 
This construction can give some new insights into the appearance of many-body ME and the quantum avalanche phenomenon in MBL \cite{Thiery2018Many, Rubio2019Many, See2022Many}.

\textit{Physical model and methods}: The coupled random matrix model can be written as 
\begin{equation}
	\mathcal{H} = \begin{pmatrix}
H_0 & 0 \\
0 & H_1
\end{pmatrix} + \frac{g}{N^{\gamma/2}} \begin{pmatrix}
0 & V \\ V^{\dagger} & 0
\end{pmatrix},
\label{eq-coupled-model}
\end{equation} 
where $H_0$ and $H_1$ are $N \times N$ diagonal random matrix belonging to Poisson Ensemble (PE) with 
$(H_{\sigma})_{ii}$ uniformly distributed in $(M_\sigma - U_\sigma/2, M_\sigma + U_\sigma/2)$, for $\sigma =0$, 1.  
The entries of $V$ are chosen to be independently distributed as $P(V_{ij}) = (1 - c)\delta(V_{ij}) + h(V_{ij})c$ \cite{NoteCentralLimit}, where $c = N^{\nu-1}$ ($ 0\leq \nu \leq 1$) controls the degree of connectance of the matrix, and $h(x) = \exp(-x^2/2)/\sqrt{2\pi}$.  
The constant $g$ is not essential when $\nu\neq 0$, thus hereafter we set $g=1$ \citep{Kravtsov2015Random}. 
We expect this model can host the ergodic, localized, and NEE phases, and tunable MEs. 
The general theory for the NEE phase and ME, and verification of their universality are the main tasks of this work. 

We will verify the predictions from the statistics of 
eigenvalues and eigenvectors. The level-spacing ratio is
defined as $ r_n = \min(s_n, s_{n+1})/\max(s_n, s_{n+1})$, where $s_n = E_{n+1} - E_n$ is the level-spacing between two adjacent energy levels.
Its mean value is taken to be $\langle r \rangle \approx 0.5307$ for the ergodic and NEE phases while $\langle r \rangle \approx 0.386$ for the localized phase \cite{Kravtsov2015Random, Khaymovich2020Fragile}. Next, the fractal dimension of wave functions is defined as $D_q(E_n, N) = -\ln(\sum_m |\langle m|\psi_n \rangle |^{2q})/[\ln(2N)(1-q)]$ with $| \psi_n\rangle$ being the eigenvectors.
In the large $N$ limit, $D_q(E_n) = \lim_{N\rightarrow\infty} D_q(E_n, N)$, distinguishing the localized ($D_q = 0$), ergodic ($D_q = 1$), and NEE ($0<D_q <1$) phases \cite{Kravtsov2015Random}. 

\textit{Theory of phase transition and phase diagram}: We first present our theory
of phase transition with only PEs in the diagonal block of Eq. \ref{eq-coupled-model}. 
When $\gamma \gg 1$ (or $g=0$), the influence of $V$ can be neglected, yielding the eigenvectors $H_{\sigma} |i\sigma\rangle = b_{i\sigma}|i\sigma\rangle$, where $|i\sigma\rangle = |\psi_{i,\sigma}\rangle$, with $b_{i,\sigma}$ being the corresponding eigenvalue.  When $\gamma$ around unity, coupling between these two matrices is relevant, 
which can be understood from Fermi's Golden Rule  as \cite{sakurai1995modern, Micklitz2022Emergence, Venturelli2023Replica}
\begin{equation}
    \Gamma(b_{i\sigma}) = 2\pi\sum_{j,\sigma'} |\langle i\sigma| T(b_{i\sigma}) |j\sigma'\rangle|^2\delta(b_{i\sigma} - b_{j\sigma'}),
    \label{eq-FGR}
\end{equation}
with $T(E)$ the transition operator. 
The quantity $\Gamma$ can be interpreted as the bandwidth of the state $|i \sigma \rangle$ perturbed by $V$, which implies that the eigenstates within the energy window of $[b_{i\sigma} - \Gamma/2, b_{i\sigma} +\Gamma/2]$ are strongly hybridized \cite{Venturelli2023Replica}.
Thus, the two transition points $\gamma_\text{ergo}$ and $\gamma_\text{AL}$, which denote the ergodic-NEE transition and NEE-localized transition points (see Fig. \ref{fig-schematic} (d)), can be characterized by the following $\beta$ functions
\begin{eqnarray}
    \beta_\text{ergo} = \frac{d\ln(\Gamma/U)}{d\ln(N)}, \quad \beta_\text{AL} = \frac{d\ln(\Gamma/\delta_\text{eff})}{d\ln(N)},
    \label{eq-beta-function}
\end{eqnarray}
with $U \propto N^{0}$ the width of the spectra for global coupling and $\delta_\text{eff}$ the effective averaged level-spacing for local coupling. 
This yields Fig. \ref{fig-schematic} (b) - (c). When $\nu = 1$, we have $\delta_\text{eff} \sim U N^{-1}$, in agreement with the averaged level-spacing of $H_0$ and $H_1$.
When $0<\nu<1$, we have $\delta_\text{eff} \sim U N^{-\nu}$ because only $N^{\nu}$ states are coupled by the coupling $V$ on average.
Physically, in the language of renormalization group theory, $\beta<0$ ($\beta>0$) means that $\Gamma/U$ or $\Gamma/\delta_\text{eff}$ are irrelevant (relevant) and flows to zero (infinite) with the increasing of $N$. 
Therefore, the condition $\beta = 0$ is marginal, which determines the value of $\gamma_\text{ergo}$ and $\gamma_\text{AL}$.  

\begin{figure}[hbtp]
\centering
\includegraphics[width=0.48\textwidth]{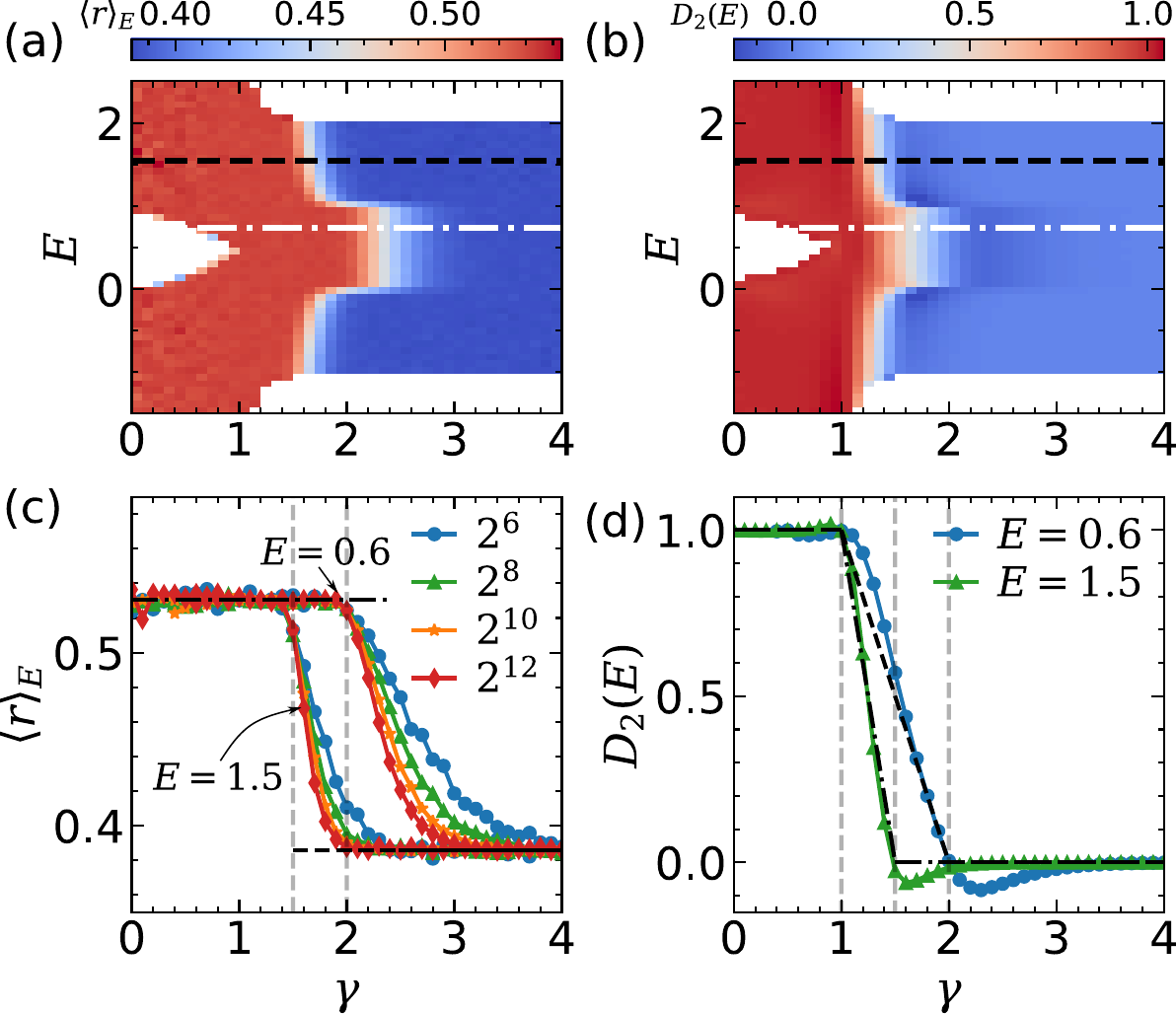}
\caption{
(a) The level-spacing ratio $\langle r\rangle_{E}$ against index $\gamma$ and energy $E$ with $N = 2^{12}$ and $\nu = 1$.
(b) The fractal dimension $D_2(E)$ against $\gamma$ and $E$. The horizontal lines in (a) and (b) denote $E=0.6$ and $E=1.5$. (c) A detail plot of $\langle r \rangle_E$ at different sizes. 
The cross points are at $\gamma = 3/2$ and $\gamma=2$ (see the two grey vertical lines) for $\beta_\text{AL}$ in the overlapped and un-overlapped regimes, respectively. 
(d) The fractal dimension $D_2(E)$ for $|E - 0.6| < \delta$
and $|E - 1.5| < \delta$, denoted as $E =0.6$ and 1.5, with $\delta = 0.05$ for $\beta_\text{ergo} = 1$.}
\label{fig-PE-PE-nu-1}
\end{figure}

One of the central observations is that the overlapped and un-overlapped spectra will exhibit totally different behaviors, as pointed out in (I). In the overlapped spectra (see Fig. \ref{fig-schematic} (a)), we have $\langle i\sigma | V |j \sigma' \rangle = V_{ij}/N^{\gamma/2} \neq 0$ for $b_{i\sigma} \sim b_{j\sigma'}$ from direct couplings. 
Taking ensemble average and energy conservation, we obtain the decay rate to the leading order (see S3 in Ref. \cite{SI})
\begin{equation}
    \Gamma_\text{ov} = 2\pi \sum_j |V_{ij}|^2 N^{-\gamma} \sim 2\pi\rho(b_{i\sigma}) N^{\nu-\gamma}.
    \label{eq-width-overlap}
\end{equation}
In contrast, in the un-overlapped spectra, the first-order process is forbidden and the second-order term is dominated, with $ T(b_{i\sigma}) \approx V(E-H_{1-\sigma})^{-1}V$, yielding
\begin{equation}
    \Gamma_\text{un} = \frac{2\pi}{N^{2\gamma}} \sum_{j \neq i} |\langle i \sigma |T(b_{i\sigma})|j\sigma\rangle|^2    
    \propto \frac{2\pi\rho(b_{i\sigma})}{N^{2(\gamma - \nu)}}.
    \label{eq-width-un-overlap}
\end{equation}
Thus the two regimes have totally different scaling behaviors with respect to $N$.
It should be noticed that we have neglected the possible higher-order terms (see S3 in Ref. \cite{SI}), which will not affect the existence of the NEE phase and MEs in our model.
Combining Eq. \ref{eq-beta-function} -  Eq. \ref{eq-width-un-overlap} lead to $\beta$ function for the overlapped spectra as
\begin{equation}
    \beta_\text{ergo} = \nu - \gamma, \quad \beta_\text{AL} = 2\nu - \gamma,
    \label{eq-beta-overlap}
\end{equation}
and that for the un-overlapped spectra as
\begin{equation}
    \beta_\text{ergo} = 2(\nu - \gamma), \quad \beta_\text{AL} = 3\nu - 2\gamma.
    \label{eq-beta-un-overlap}
\end{equation}
$\beta = 0$ determines the phase boundaries, thus we have $\gamma_\text{ergo} = \nu$, $\gamma_\text{AL} = 2\nu$ for the overlapped spectra and $\gamma_\text{ergo} = \nu$, $\gamma_\text{AL} = 3\nu/2$ for the un-overlapped spectra, yielding the phase diagram in Fig. \ref{fig-schematic} (d) and (e).
So the MEs in the context of correlation of energy levels exist when $3\nu/2<\gamma<2\nu$ and the MEs in the context of wave function exist when $\nu<\gamma<2\nu$.
It is found that $\gamma_\text{ergo} = \nu$ for both overlapped and un-overlapped spectra, and when $\gamma < \nu$ (for $\beta_\text{ergo} >0$), all states are strongly coupled. from strong global coupling. 

The above analysis is a generalization of the approach presented in RP model \citep{Khaymovich2020Fragile, Biroli2021Levy} based on the divergence of $\sum_j |H_{ij}|$ and 
$\sum_j |H_{ij}|^2$, which also yield Eq. \ref{eq-beta-overlap} \cite{RPnoteofExponent}. 
Our approach has a somewhat much clearer physical meaning, and meanwhile can be generalized to block coupled models. 
We find that in the regime $\gamma \in (\nu, 3\nu/2)$ (un-overlapped regime) and $\gamma \in (\nu, 2\nu)$ (overlapped regime), the states can be coupled only through consequential coupling between neighboring energy states and form mini-bands in the local spectrum 
\cite{Khaymovich2020Fragile, Venturelli2023Replica}.
The above results are significant because, in the sparse matrix limit of $\nu = 0$, we have $\gamma_\text{ergo} = \gamma_\text{AL} = 0$, which corresponds to the many-body disordered spin models and Hubbard models widely studied in the literature \cite{Chanda2020Many-body, Deng2017Many-body, Wei2019Investigating} and sparse matrix researched by Mirlin \textit{et. al.} \cite{mirlin1991universality,fyodorov1991localization}. In this condition, $g$ is marginal, and a finite $g_c$ is required for the localization-delocalization transition. 

\begin{figure}[hbtp]
\centering
\includegraphics[width=0.46\textwidth]{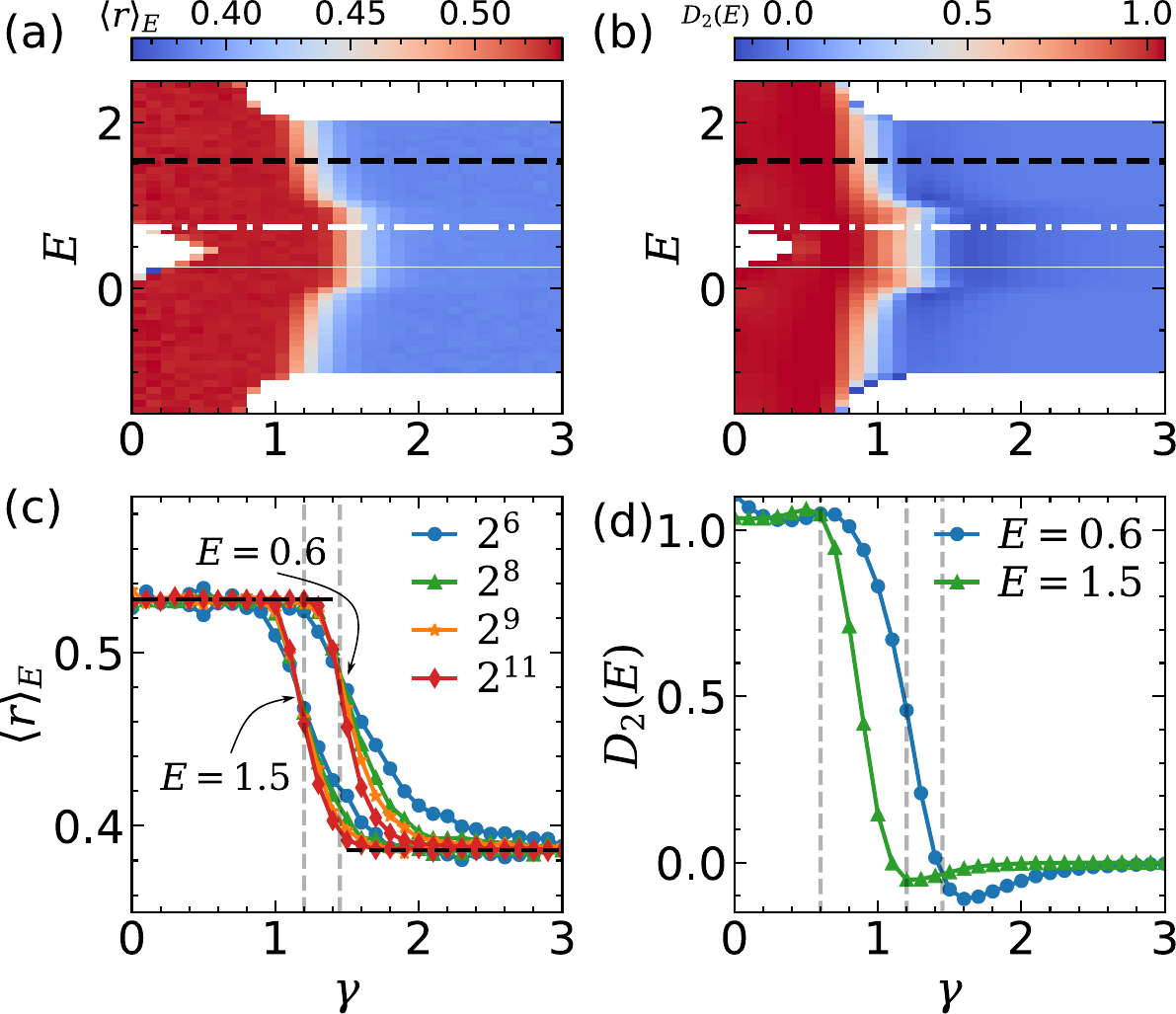}
\caption{The same as that in Fig. \ref{fig-PE-PE-nu-1} but with $\nu = 0.6$.
All the major features of phase transition in Fig. \ref{fig-PE-PE-nu-1} are maintained, except that in (c) the two cross points happen at $\gamma_\text{AL} = 1.39$ in the overlapped spectra and $\gamma_\text{AL} = 1.16$ in the un-overlapped spectra; while from the theoretical prediction it should be $1.2$ and $0.9$, respectively. 
This may come from the finite size effect (see S4 of Ref. \cite{SI}, from which we find the apparent cross points $\gamma_c$ shift towards the theoretic expectation with the increase of $N$) or the neglect of the higher-order terms in perturbation theory. 
In (d) from the fractal dimension $D_2(E)$ for $E=0.6$ and $E=1.5$, we still have $\beta_\text{ergo} = \nu = 0.6$; see Fig. \ref{fig-PE-PE-nu-1}. }
\label{fig-PE-PE-nu-06}
\end{figure}

It is extremely intriguing to discuss the properties of the NEE phase, whose fractal dimension is not an integer.
The support set of the eigenstates is simply given by the bandwidth $\Gamma$ divided by the averaged level spacing $\delta_\text{eff} \sim N^{-\nu}$, which is a result of the local coupling.
Thus the non-zero component of the new eigenstate scales as $\mathcal{N} \sim \Gamma/ N^{-\nu}$ and it can be written as $ |\psi_{i,\sigma}^{\prime}\rangle = \sum_{j,\sigma', |b_{i\sigma} - b_{j\sigma'}|< \Gamma/2} c_{j,\sigma'} |j\sigma'\rangle$, with $c_{j,\sigma'} \sim (\Gamma N^{\nu})^{-1/2}$ for normalization \cite{Venturelli2023Replica}.
The fractal dimension of these wave functions is given by $D_{q} = \lim_{N\rightarrow\infty}\ln((\Gamma N^{\nu})^{1-q} )/(\ln(2N)(1-q))$. Using the expression of $\Gamma$ in Eq. \ref{eq-FGR} will yield 
\begin{equation}
    D_q^\text{ov} = 2 \nu-\gamma, \quad D_q^\text{un} = 3\nu -2\gamma,
    \label{eq-fd}
\end{equation}
for $\gamma_\text{ergo}<\gamma<\gamma_\text{AL}$.
Thus, the NEE phase, without $q$-dependence, is fractal but not multifractal, setting it apart from from that in the RP ensemble with a heavy-tail distribution \citep{Khaymovich2020Fragile, Biroli2021Levy}. 
With heavy-tail distributed elements, the existence of ME has also been reported \cite{tarquini2016level,aggarwal2019goe}. However, the underlying mechanism is different. 

With this theoretical analysis, it is essential to verify the above results numerically. 
We use $M_0 = 0$, $M_1 = 1$ and $U_0 = U_1 = 2$ in Eq. \ref{eq-coupled-model}. Thus, the spectra of $H_0$ and $H_1$ overlap within energy regime $[0,1]$.
The results with $\nu = 1$ are shown in Fig. \ref{fig-PE-PE-nu-1}, which agree excellently with the theoretical prediction in Eq. \ref{eq-beta-overlap} and Eq. \ref{eq-beta-un-overlap}. 
The results with $\nu = 0.6$ are presented in Fig. \ref{fig-PE-PE-nu-06}. 
We find that the numerical results indicate $\gamma_\text{ergo} = 1.16$ and $\gamma_\text{AL} = 1.39$, being slightly different from the prediction based on perturbation theory.
This should come from the finite size effect and we show in S4 \cite{SI} that these two limits can be approached in the large $N$ limit.

\begin{figure}[hbtp]
\centering
\includegraphics[width=0.45\textwidth]{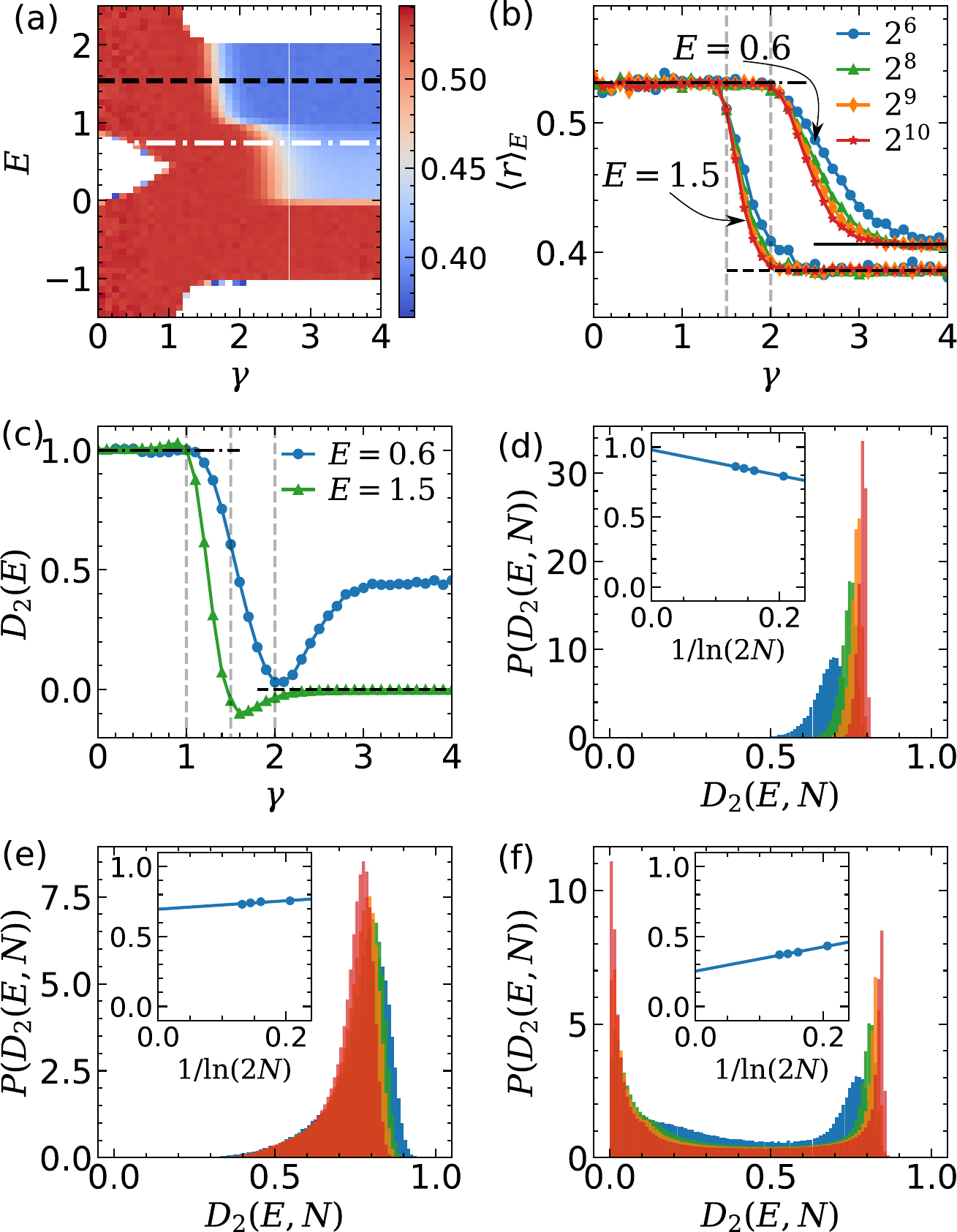}
\caption{
(a) The averaged level-spacing ratio $\langle r\rangle_{E}$ against index $\gamma$ and energy $E$.
The horizontal lines denote $E=0.6$ and $E=1.5$.
(b) A detailed plot of $\langle r \rangle_E$ at different sizes. 
The cross points are at $\gamma = 3/2$ and $\gamma=2$, corresponding to the grey vertical lines.
(c) The fractal dimension $D_2(E)$ for $E=0.6$ and $E=1.5$.
(d)-(f) The distribution function for finite-size $D_2(E,N)$ with $0<E<1$ at sizes $N=2^6$ (blue), $N=2^8$ (green), $N=2^9$ (orange), $N=2^{10}$ (red).
We use $\gamma = 0.5$ in (d), $\gamma = 1.5$ in (e), and $\gamma = 2.5$ in (f).
The insets show the averaged $D_2(E,N)$ of states over $E \in [0,1]$ against $1/\ln(2N)$.
The solid lines are fitted according to $\langle D_2(E, N)\rangle_E = A/\ln(2N) + D_2(E)$.
}
\label{fig-PE-GOE}
\end{figure}

\textit{Universality of this mechanism}: 
Since the block structure, instead of the symmetry, is most essential for our model, we expect our results to be also applicable when $H_0$, $H_1$, and $V$ are replaced by other random matrices. 
However, the transition point and the associated $\beta$ functions may be affected by correlations in $V$.
To this end, we set $H_0$ as PE whose diagonal elements are distributed in the interval $[0,2]$ and $H_1$ as Gaussian orthogonal ensemble (GOE) whose elements are independent Gaussian distributed with mean zero and variance $1/\sqrt{2N}$ (see other choices of $H_0$ and $H_1$ in S5 of Ref. \cite{SI}).
By choosing these two ensembles, their spectra are overlapped in the interval $[0,1]$. Then we set $\nu = 1$ for the off-diagonal block and expect $\gamma_\text{ergo} = 1$ and $\gamma_\text{AL} = 2$.
When $\gamma > \gamma_{\text{AL}}$, the coupling is irrelevant, yielding a coexistence phase of the ergodic phase and localized phase in the overlapped spectra with some different statistics\cite{Giraud2022Probing,Berry1984Semiclassical}. 
In the un-overlapped spectra, the states are either localized or ergodic depending on their energy. 
When $\gamma < \gamma_\text{ergo}$, the off-diagonal coupling becomes relevent with the increasing of $N$, yielding the ergodic phase for the whole spectra. 
The most intriguing physics happens in the intermediate region of $\gamma$, the coupling between $H_0$ and $H_1$ induces competition of ergodic and localized states, leaving their fate undetermined.
We perform a calculation based on the exact diagonalization method to examine the phase diagram, which is presented in Fig. \ref{fig-PE-GOE}. 
The level-spacing ratio $\langle r \rangle_E$ exhibits two different cross points for overlapped and un-overlapped spectra, which is the same as the results in Fig. \ref{fig-PE-PE-nu-1}.
However, $\langle r \rangle_E \neq 0.386$ and $0<D_2(E)<1$ in the overlapped spectra when $\gamma>2$, corresponds to a coexistence phase of ergodic and localized states \citep{Berry1984Semiclassical,Giraud2022Probing}.
This is further verified by the distribution of finite-size $D_2(E, N)$ as shown in Fig. \ref{fig-PE-GOE} (d), (e), and (f).
These results indicate that there is only one peak of $P(D_2(E, N))$ for the ergodic and NEE phases while the distribution of $P(D_2(E, N))$ of the coexistence phase exhibits two peaks as shown in Fig. \ref{fig-PE-GOE} (f).

\begin{figure}[hbtp]
\centering
\includegraphics[width=0.48\textwidth]{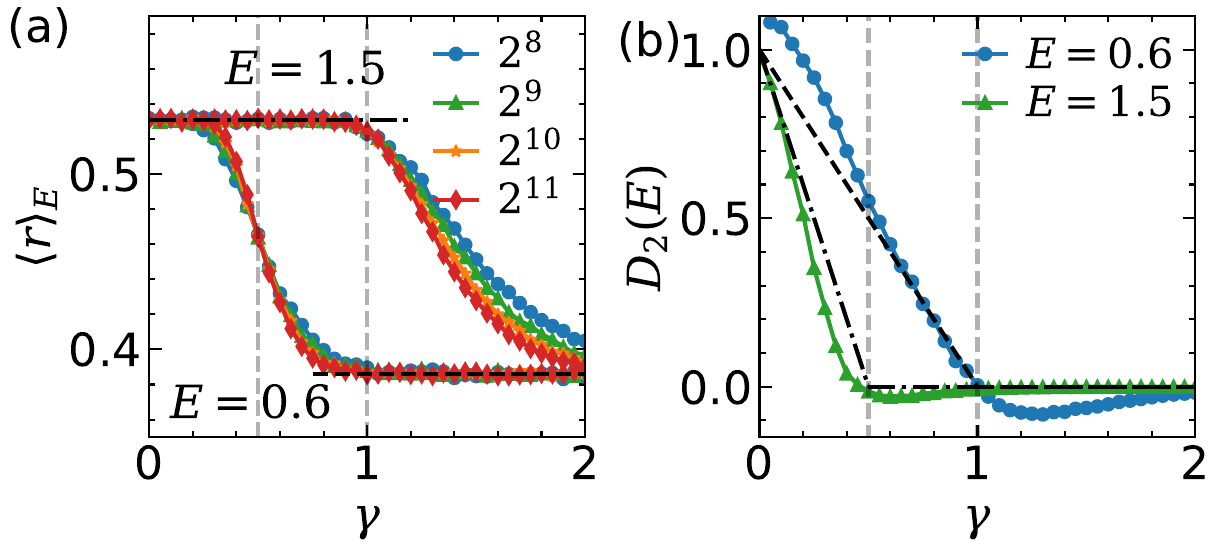}
\caption{
(a) The averaged level-spacing ratio $\langle r\rangle_{E}$ against index $\gamma$ within energy windows $|E - 1.5| \leq 0.01$ and $|E -0.6| \leq 0.01$ with $V$ being COE.
The cross points are at $\gamma = 1/2$ and $\gamma=1$, corresponding to the grey vertical lines.
(b) The fractal dimension $D_2(E)$ around $E=0.6$ (overlapped) and $E=1.5$ (un-overlapped), showing 
$\gamma_\text{AL} = 1$ (overlapped) and $\gamma_\text{AL} = 1/2$ (un-overlapped).}
\label{fig-PE-PE-COE}
\end{figure}

Another important extension of this theory is replacing the Gaussian distributed $V$ with a  circular orthogonal ensemble (COE), with $V^{T}V = 1$ \cite{dyson1962brownian, Forrester2010LogGas}. 
This is expected if we set $H_0$ and $H_1$ as random matrices, which can be diagonalized using two unitary matrices $U_1$ and $U_2$. Then the off-diagonal matrix should be $V = U_1^\dagger T U_2$, which can be a COE when $T = \mathbb{I}$ is unity. 
This matrix, unlike the Gaussian counterpart, is correlated, and it is expected that $\langle (V_{ij})^2\rangle \sim N^{-1}$, since $\text{Tr} (V^\dagger V) = N$.
As compared with the Gaussian distributed $V$, we have to define the effective parameters as $\gamma_\text{eff} = 1 + \gamma$, following the above procedure. It yields $\gamma_\text{ergo} = 0$, $\gamma_\text{AL} = 1$ for the overlapped spectra and $\gamma_\text{ergo} = 0$, $\gamma_\text{AL} = 1/2$ for the un-overlapped spectra.
The verification of the $\gamma_\text{AL}$ is presented
in Fig. \ref{fig-PE-PE-COE}, setting $V$ to be a COE, showing
$\gamma_\text{AL} = 1/2$ and $1$ in these two different energy regimes. 
Here, we also chose the energy regime $[0,1]$ as the overlapped spectra.

\textit{Conclusion and Remarks}: We present a new class of random matrices based on two coupled matrices and show that the overlapped and un-overlapped spectra have totally different scaling behaviors, which can be used to construct tunable MEs. 
This theory applies to dense, sparse, and even corrected random matrices. 
Our study opens several intriguing research directions. 
Firstly, it provides a universal framework for the realization of MEs in random matrices and even realistic models, which offer exceptional opportunities for the experimental detection of many-body MEs. 
Secondly, our approach could be applied to the possible MEs in non-hermitian random models with overlapped spectra in the complex plane \citep{Tomasi2022Non-Hermitian, Hamazaki2019Non-hermitian, Gong2018Topological, Ashida2021Nonermite}. 
Finally, it can be generalized to multi-block random matrices, which are used for MBL in realistic models. Applications of our model may include ME and localization-delocalization transitions in 
many-body models \citep{Thiery2018Many, Rubio2019Many, See2022Many}, chiral symmetry breaking of the Dirac operator in QCD \cite{Stephanov1996RandomMatrixModel, Verbaarschot1994Spectrum, Verbaarschot2000Random} and 
stability of large ecological communities \cite{May1972Will, Servan2018Coexistence, Grilli2017Feasibility,EcosystemnoteNote}. This model itself is also of general interest in pure mathematics in context of RMT \cite{Forrester2010LogGas}.

\textit{Acknowledgments}: 
This work is supported by the National Natural Science Foundation of China (NSFC) with No. 11774328, and Innovation Program for Quantum Science and Technology (No. 2021ZD0301200 and No. 2021ZD0301500).

\bibliography{ref} 

\begin{thebibliography}{77}%
\makeatletter
\providecommand \@ifxundefined [1]{%
 \@ifx{#1\undefined}
}%
\providecommand \@ifnum [1]{%
 \ifnum #1\expandafter \@firstoftwo
 \else \expandafter \@secondoftwo
 \fi
}%
\providecommand \@ifx [1]{%
 \ifx #1\expandafter \@firstoftwo
 \else \expandafter \@secondoftwo
 \fi
}%
\providecommand \natexlab [1]{#1}%
\providecommand \enquote  [1]{``#1''}%
\providecommand \bibnamefont  [1]{#1}%
\providecommand \bibfnamefont [1]{#1}%
\providecommand \citenamefont [1]{#1}%
\providecommand \href@noop [0]{\@secondoftwo}%
\providecommand \href [0]{\begingroup \@sanitize@url \@href}%
\providecommand \@href[1]{\@@startlink{#1}\@@href}%
\providecommand \@@href[1]{\endgroup#1\@@endlink}%
\providecommand \@sanitize@url [0]{\catcode `\\12\catcode `\$12\catcode
  `\&12\catcode `\#12\catcode `\^12\catcode `\_12\catcode `\%12\relax}%
\providecommand \@@startlink[1]{}%
\providecommand \@@endlink[0]{}%
\providecommand \url  [0]{\begingroup\@sanitize@url \@url }%
\providecommand \@url [1]{\endgroup\@href {#1}{\urlprefix }}%
\providecommand \urlprefix  [0]{URL }%
\providecommand \Eprint [0]{\href }%
\providecommand \doibase [0]{http://dx.doi.org/}%
\providecommand \selectlanguage [0]{\@gobble}%
\providecommand \bibinfo  [0]{\@secondoftwo}%
\providecommand \bibfield  [0]{\@secondoftwo}%
\providecommand \translation [1]{[#1]}%
\providecommand \BibitemOpen [0]{}%
\providecommand \bibitemStop [0]{}%
\providecommand \bibitemNoStop [0]{.\EOS\space}%
\providecommand \EOS [0]{\spacefactor3000\relax}%
\providecommand \BibitemShut  [1]{\csname bibitem#1\endcsname}%
\let\auto@bib@innerbib\@empty
\bibitem [{\citenamefont {Forrester}(2010)}]{Forrester2010LogGas}%
  \BibitemOpen
  \bibfield  {author} {\bibinfo {author} {\bibfnamefont {Peter~J.}\
  \bibnamefont {Forrester}},\ }\href {\doibase doi:10.1515/9781400835416}
  {\emph {\bibinfo {title} {Log-Gases and Random Matrices (LMS-34)}}}\
  (\bibinfo  {publisher} {Princeton University Press},\ \bibinfo {address}
  {Princeton},\ \bibinfo {year} {2010})\BibitemShut {NoStop}%
\bibitem [{\citenamefont {Brody}\ \emph {et~al.}(1981)\citenamefont {Brody},
  \citenamefont {Flores}, \citenamefont {French}, \citenamefont {Mello},
  \citenamefont {Pandey},\ and\ \citenamefont {Wong}}]{Brody1981Randommatrix}%
  \BibitemOpen
  \bibfield  {author} {\bibinfo {author} {\bibfnamefont {T.~A.}\ \bibnamefont
  {Brody}}, \bibinfo {author} {\bibfnamefont {J.}~\bibnamefont {Flores}},
  \bibinfo {author} {\bibfnamefont {J.~B.}\ \bibnamefont {French}}, \bibinfo
  {author} {\bibfnamefont {P.~A.}\ \bibnamefont {Mello}}, \bibinfo {author}
  {\bibfnamefont {A.}~\bibnamefont {Pandey}}, \ and\ \bibinfo {author}
  {\bibfnamefont {S.~S.~M.}\ \bibnamefont {Wong}},\ }\bibfield  {title}
  {\enquote {\bibinfo {title} {Random-matrix physics: spectrum and strength
  fluctuations},}\ }\href {\doibase 10.1103/RevModPhys.53.385} {\bibfield
  {journal} {\bibinfo  {journal} {Rev. Mod. Phys.}\ }\textbf {\bibinfo {volume}
  {53}},\ \bibinfo {pages} {385--479} (\bibinfo {year} {1981})}\BibitemShut
  {NoStop}%
\bibitem [{\citenamefont {Mehta}(2004)}]{Mehta2004Random}%
  \BibitemOpen
  \bibfield  {author} {\bibinfo {author} {\bibfnamefont {Madan~Lal}\
  \bibnamefont {Mehta}},\ }\href@noop {} {\emph {\bibinfo {title} {Random
  matrices}}}\ (\bibinfo  {publisher} {Elsevier},\ \bibinfo {year}
  {2004})\BibitemShut {NoStop}%
\bibitem [{\citenamefont {Livan}\ \emph {et~al.}(2018)\citenamefont {Livan},
  \citenamefont {Novaes},\ and\ \citenamefont {Vivo}}]{Livan2018Introduction}%
  \BibitemOpen
  \bibfield  {author} {\bibinfo {author} {\bibfnamefont {Giacomo}\ \bibnamefont
  {Livan}}, \bibinfo {author} {\bibfnamefont {Marcel}\ \bibnamefont {Novaes}},
  \ and\ \bibinfo {author} {\bibfnamefont {Pierpaolo}\ \bibnamefont {Vivo}},\
  }\bibfield  {title} {\enquote {\bibinfo {title} {Introduction to random
  matrices theory and practice},}\ }\href@noop {} {\bibfield  {journal}
  {\bibinfo  {journal} {Monograph Award}\ }\textbf {\bibinfo {volume} {63}}
  (\bibinfo {year} {2018})}\BibitemShut {NoStop}%
\bibitem [{\citenamefont {Wigner}(1955)}]{Wigner1955Characteristic}%
  \BibitemOpen
  \bibfield  {author} {\bibinfo {author} {\bibfnamefont {Eugene~P.}\
  \bibnamefont {Wigner}},\ }\bibfield  {title} {\enquote {\bibinfo {title}
  {Characteristic {Vectors} of {Bordered} {Matrices} {With} {Infinite}
  {Dimensions}},}\ }\href {\doibase 10.2307/1970079} {\bibfield  {journal}
  {\bibinfo  {journal} {Annals of Mathematics}\ }\textbf {\bibinfo {volume}
  {62}},\ \bibinfo {pages} {548} (\bibinfo {year} {1955})}\BibitemShut
  {NoStop}%
\bibitem [{\citenamefont
  {Dyson}(1962{\natexlab{a}})}]{dyson1962brownianmotion}%
  \BibitemOpen
  \bibfield  {author} {\bibinfo {author} {\bibfnamefont {Freeman~J.}\
  \bibnamefont {Dyson}},\ }\bibfield  {title} {\enquote {\bibinfo {title} {A
  {Brownian}‐{Motion} {Model} for the {Eigenvalues} of a {Random}
  {Matrix}},}\ }\href {\doibase 10.1063/1.1703862} {\bibfield  {journal}
  {\bibinfo  {journal} {J. Math. Phys.}\ }\textbf {\bibinfo {volume} {3}},\
  \bibinfo {pages} {1191} (\bibinfo {year} {1962}{\natexlab{a}})}\BibitemShut
  {NoStop}%
\bibitem [{\citenamefont {Bohigas}\ \emph {et~al.}(1984)\citenamefont
  {Bohigas}, \citenamefont {Giannoni},\ and\ \citenamefont
  {Schmit}}]{Bohigas1982Characterization}%
  \BibitemOpen
  \bibfield  {author} {\bibinfo {author} {\bibfnamefont {O.}~\bibnamefont
  {Bohigas}}, \bibinfo {author} {\bibfnamefont {M.~J.}\ \bibnamefont
  {Giannoni}}, \ and\ \bibinfo {author} {\bibfnamefont {C.}~\bibnamefont
  {Schmit}},\ }\bibfield  {title} {\enquote {\bibinfo {title} {Characterization
  of chaotic quantum spectra and universality of level fluctuation laws},}\
  }\href {\doibase 10.1103/PhysRevLett.52.1} {\bibfield  {journal} {\bibinfo
  {journal} {Phys. Rev. Lett.}\ }\textbf {\bibinfo {volume} {52}},\ \bibinfo
  {pages} {1} (\bibinfo {year} {1984})}\BibitemShut {NoStop}%
\bibitem [{\citenamefont {Berry}\ \emph {et~al.}(1977)\citenamefont {Berry},
  \citenamefont {Tabor},\ and\ \citenamefont {Ziman}}]{Berry1977Level}%
  \BibitemOpen
  \bibfield  {author} {\bibinfo {author} {\bibfnamefont {Michael~Victor}\
  \bibnamefont {Berry}}, \bibinfo {author} {\bibfnamefont {M.}~\bibnamefont
  {Tabor}}, \ and\ \bibinfo {author} {\bibfnamefont {John~Michael}\
  \bibnamefont {Ziman}},\ }\bibfield  {title} {\enquote {\bibinfo {title}
  {Level clustering in the regular spectrum},}\ }\href {\doibase
  10.1098/rspa.1977.0140} {\bibfield  {journal} {\bibinfo  {journal} {Proc. R.
  Soc. Lond. A}\ }\textbf {\bibinfo {volume} {356}},\ \bibinfo {pages} {375}
  (\bibinfo {year} {1977})}\BibitemShut {NoStop}%
\bibitem [{\citenamefont {Kos}\ \emph {et~al.}(2018)\citenamefont {Kos},
  \citenamefont {Ljubotina},\ and\ \citenamefont {Prosen}}]{Kos2018Many-body}%
  \BibitemOpen
  \bibfield  {author} {\bibinfo {author} {\bibfnamefont {Pavel}\ \bibnamefont
  {Kos}}, \bibinfo {author} {\bibfnamefont {Marko}\ \bibnamefont {Ljubotina}},
  \ and\ \bibinfo {author} {\bibfnamefont {Tomaž}\ \bibnamefont {Prosen}},\
  }\bibfield  {title} {\enquote {\bibinfo {title} {Many-{Body} {Quantum}
  {Chaos}: {Analytic} {Connection} to {Random} {Matrix} {Theory}},}\ }\href
  {\doibase 10.1103/PhysRevX.8.021062} {\bibfield  {journal} {\bibinfo
  {journal} {Phys. Rev. X}\ }\textbf {\bibinfo {volume} {8}},\ \bibinfo {pages}
  {021062} (\bibinfo {year} {2018})}\BibitemShut {NoStop}%
\bibitem [{\citenamefont {Mirlin}(2000)}]{Mirlin2000Statistics}%
  \BibitemOpen
  \bibfield  {author} {\bibinfo {author} {\bibfnamefont {Alexander~D.}\
  \bibnamefont {Mirlin}},\ }\bibfield  {title} {\enquote {\bibinfo {title}
  {Statistics of energy levels and eigenfunctions in disordered systems},}\
  }\href {\doibase 10.1016/S0370-1573(99)00091-5} {\bibfield  {journal}
  {\bibinfo  {journal} {Physics Reports}\ }\textbf {\bibinfo {volume} {326}},\
  \bibinfo {pages} {259--382} (\bibinfo {year} {2000})}\BibitemShut {NoStop}%
\bibitem [{\citenamefont {Goda}\ \emph {et~al.}(2006)\citenamefont {Goda},
  \citenamefont {Nishino},\ and\ \citenamefont {Matsuda}}]{Goda2006InverseAL}%
  \BibitemOpen
  \bibfield  {author} {\bibinfo {author} {\bibfnamefont {Masaki}\ \bibnamefont
  {Goda}}, \bibinfo {author} {\bibfnamefont {Shinya}\ \bibnamefont {Nishino}},
  \ and\ \bibinfo {author} {\bibfnamefont {Hiroki}\ \bibnamefont {Matsuda}},\
  }\bibfield  {title} {\enquote {\bibinfo {title} {Inverse anderson transition
  caused by flatbands},}\ }\href {\doibase 10.1103/PhysRevLett.96.126401}
  {\bibfield  {journal} {\bibinfo  {journal} {Phys. Rev. Lett.}\ }\textbf
  {\bibinfo {volume} {96}},\ \bibinfo {pages} {126401} (\bibinfo {year}
  {2006})}\BibitemShut {NoStop}%
\bibitem [{\citenamefont {Pal}\ and\ \citenamefont
  {Huse}(2010)}]{Pal2010Many-body}%
  \BibitemOpen
  \bibfield  {author} {\bibinfo {author} {\bibfnamefont {Arijeet}\ \bibnamefont
  {Pal}}\ and\ \bibinfo {author} {\bibfnamefont {David~A.}\ \bibnamefont
  {Huse}},\ }\bibfield  {title} {\enquote {\bibinfo {title} {Many-body
  localization phase transition},}\ }\href {\doibase
  10.1103/PhysRevB.82.174411} {\bibfield  {journal} {\bibinfo  {journal} {Phys.
  Rev. B}\ }\textbf {\bibinfo {volume} {82}},\ \bibinfo {pages} {174411}
  (\bibinfo {year} {2010})}\BibitemShut {NoStop}%
\bibitem [{\citenamefont {Nandkishore}\ and\ \citenamefont
  {Huse}(2015)}]{Nandkishore2015Many-body}%
  \BibitemOpen
  \bibfield  {author} {\bibinfo {author} {\bibfnamefont {Rahul}\ \bibnamefont
  {Nandkishore}}\ and\ \bibinfo {author} {\bibfnamefont {David~A.}\
  \bibnamefont {Huse}},\ }\bibfield  {title} {\enquote {\bibinfo {title}
  {Many-{Body} {Localization} and {Thermalization} in {Quantum} {Statistical}
  {Mechanics}},}\ }\href {\doibase 10.1146/annurev-conmatphys-031214-014726}
  {\bibfield  {journal} {\bibinfo  {journal} {Annual Review of Condensed Matter
  Physics}\ }\textbf {\bibinfo {volume} {6}},\ \bibinfo {pages} {15} (\bibinfo
  {year} {2015})}\BibitemShut {NoStop}%
\bibitem [{\citenamefont {Abanin}\ \emph {et~al.}(2019)\citenamefont {Abanin},
  \citenamefont {Altman}, \citenamefont {Bloch},\ and\ \citenamefont
  {Serbyn}}]{Abanin2019Colloquium}%
  \BibitemOpen
  \bibfield  {author} {\bibinfo {author} {\bibfnamefont {Dmitry~A.}\
  \bibnamefont {Abanin}}, \bibinfo {author} {\bibfnamefont {Ehud}\ \bibnamefont
  {Altman}}, \bibinfo {author} {\bibfnamefont {Immanuel}\ \bibnamefont
  {Bloch}}, \ and\ \bibinfo {author} {\bibfnamefont {Maksym}\ \bibnamefont
  {Serbyn}},\ }\bibfield  {title} {\enquote {\bibinfo {title} {Colloquium:
  {Many}-body localization, thermalization, and entanglement},}\ }\href
  {\doibase 10.1103/RevModPhys.91.021001} {\bibfield  {journal} {\bibinfo
  {journal} {Rev. Mod. Phys.}\ }\textbf {\bibinfo {volume} {91}},\ \bibinfo
  {pages} {021001} (\bibinfo {year} {2019})}\BibitemShut {NoStop}%
\bibitem [{\citenamefont {Alet}\ and\ \citenamefont
  {Laflorencie}(2018)}]{Alet2018Many-body}%
  \BibitemOpen
  \bibfield  {author} {\bibinfo {author} {\bibfnamefont {Fabien}\ \bibnamefont
  {Alet}}\ and\ \bibinfo {author} {\bibfnamefont {Nicolas}\ \bibnamefont
  {Laflorencie}},\ }\bibfield  {title} {\enquote {\bibinfo {title} {Many-body
  localization: an introduction and selected topics},}\ }\href {\doibase
  10.1016/j.crhy.2018.03.003} {\bibfield  {journal} {\bibinfo  {journal}
  {Comptes Rendus Physique}\ }\textbf {\bibinfo {volume} {19}},\ \bibinfo
  {pages} {498} (\bibinfo {year} {2018})}\BibitemShut {NoStop}%
\bibitem [{\citenamefont {Imbrie}(2016)}]{Imbrie2016Diagonalization}%
  \BibitemOpen
  \bibfield  {author} {\bibinfo {author} {\bibfnamefont {John~Z.}\ \bibnamefont
  {Imbrie}},\ }\bibfield  {title} {\enquote {\bibinfo {title} {Diagonalization
  and {Many}-{Body} {Localization} for a {Disordered} {Quantum} {Spin}
  {Chain}},}\ }\href {\doibase 10.1103/PhysRevLett.117.027201} {\bibfield
  {journal} {\bibinfo  {journal} {Phys. Rev. Lett.}\ }\textbf {\bibinfo
  {volume} {117}},\ \bibinfo {pages} {027201} (\bibinfo {year}
  {2016})}\BibitemShut {NoStop}%
\bibitem [{\citenamefont {Stephanov}(1996)}]{Stephanov1996RandomMatrixModel}%
  \BibitemOpen
  \bibfield  {author} {\bibinfo {author} {\bibfnamefont {M.~A.}\ \bibnamefont
  {Stephanov}},\ }\bibfield  {title} {\enquote {\bibinfo {title} {Random matrix
  model of qcd at finite density and the nature of the quenched limit},}\
  }\href {\doibase 10.1103/PhysRevLett.76.4472} {\bibfield  {journal} {\bibinfo
   {journal} {Phys. Rev. Lett.}\ }\textbf {\bibinfo {volume} {76}},\ \bibinfo
  {pages} {4472--4475} (\bibinfo {year} {1996})}\BibitemShut {NoStop}%
\bibitem [{\citenamefont {Verbaarschot}(1994)}]{Verbaarschot1994Spectrum}%
  \BibitemOpen
  \bibfield  {author} {\bibinfo {author} {\bibfnamefont {Jacobus}\ \bibnamefont
  {Verbaarschot}},\ }\bibfield  {title} {\enquote {\bibinfo {title} {Spectrum
  of the qcd dirac operator and chiral random matrix theory},}\ }\href
  {\doibase 10.1103/PhysRevLett.72.2531} {\bibfield  {journal} {\bibinfo
  {journal} {Phys. Rev. Lett.}\ }\textbf {\bibinfo {volume} {72}},\ \bibinfo
  {pages} {2531--2533} (\bibinfo {year} {1994})}\BibitemShut {NoStop}%
\bibitem [{\citenamefont {Kieburg}\ \emph {et~al.}(2013)\citenamefont
  {Kieburg}, \citenamefont {Verbaarschot},\ and\ \citenamefont
  {Zafeiropoulos}}]{Kieburg2013Spectral}%
  \BibitemOpen
  \bibfield  {author} {\bibinfo {author} {\bibfnamefont {Mario}\ \bibnamefont
  {Kieburg}}, \bibinfo {author} {\bibfnamefont {Jacobus J.~M.}\ \bibnamefont
  {Verbaarschot}}, \ and\ \bibinfo {author} {\bibfnamefont {Savvas}\
  \bibnamefont {Zafeiropoulos}},\ }\bibfield  {title} {\enquote {\bibinfo
  {title} {Spectral properties of the wilson-dirac operator and random matrix
  theory},}\ }\href {\doibase 10.1103/PhysRevD.88.094502} {\bibfield  {journal}
  {\bibinfo  {journal} {Phys. Rev. D}\ }\textbf {\bibinfo {volume} {88}},\
  \bibinfo {pages} {094502} (\bibinfo {year} {2013})}\BibitemShut {NoStop}%
\bibitem [{\citenamefont {Verbaarschot}\ and\ \citenamefont
  {Wettig}(2000)}]{Verbaarschot2000Random}%
  \BibitemOpen
  \bibfield  {author} {\bibinfo {author} {\bibfnamefont {J.~J.~M.}\
  \bibnamefont {Verbaarschot}}\ and\ \bibinfo {author} {\bibfnamefont
  {T.}~\bibnamefont {Wettig}},\ }\bibfield  {title} {\enquote {\bibinfo {title}
  {Random matrix theory and chiral symmetry in qcd},}\ }\href {\doibase
  https://doi.org/10.1146/annurev.nucl.50.1.343} {\bibfield  {journal}
  {\bibinfo  {journal} {Annual Review of Nuclear and Particle Science}\
  }\textbf {\bibinfo {volume} {50}},\ \bibinfo {pages} {343 -- 410} (\bibinfo
  {year} {2000})}\BibitemShut {NoStop}%
\bibitem [{\citenamefont {May}(1972)}]{May1972Will}%
  \BibitemOpen
  \bibfield  {author} {\bibinfo {author} {\bibfnamefont {Robert~M.}\
  \bibnamefont {May}},\ }\bibfield  {title} {\enquote {\bibinfo {title} {Will a
  large complex system be stable?}}\ }\href
  {https://www.nature.com/articles/238413a0/} {\bibfield  {journal} {\bibinfo
  {journal} {Nature}\ }\textbf {\bibinfo {volume} {238}},\ \bibinfo {pages}
  {413–414} (\bibinfo {year} {1972})}\BibitemShut {NoStop}%
\bibitem [{\citenamefont {Serván}\ \emph {et~al.}(2018)\citenamefont
  {Serván}, \citenamefont {Capitán}, \citenamefont {Grilli}, \citenamefont
  {Morrison},\ and\ \citenamefont {Allesina}}]{Servan2018Coexistence}%
  \BibitemOpen
  \bibfield  {author} {\bibinfo {author} {\bibfnamefont {Carlos~A.}\
  \bibnamefont {Serván}}, \bibinfo {author} {\bibfnamefont {José~A.}\
  \bibnamefont {Capitán}}, \bibinfo {author} {\bibfnamefont {Jacopo}\
  \bibnamefont {Grilli}}, \bibinfo {author} {\bibfnamefont {Kent~E.}\
  \bibnamefont {Morrison}}, \ and\ \bibinfo {author} {\bibfnamefont {Stefano}\
  \bibnamefont {Allesina}},\ }\bibfield  {title} {\enquote {\bibinfo {title}
  {Coexistence of many species in random ecosystems},}\ }\href {\doibase
  https://doi.org/10.1038/s41559-018-0603-6} {\bibfield  {journal} {\bibinfo
  {journal} {Nature Ecology \& Evolution}\ }\textbf {\bibinfo {volume} {2}},\
  \bibinfo {pages} {1237–1242} (\bibinfo {year} {2018})}\BibitemShut
  {NoStop}%
\bibitem [{\citenamefont {Grilli}\ \emph {et~al.}(2017)\citenamefont {Grilli},
  \citenamefont {Adorisio}, \citenamefont {Suweis}, \citenamefont {Barabás},
  \citenamefont {Banavar}, \citenamefont {Allesina},\ and\ \citenamefont
  {Maritan}}]{Grilli2017Feasibility}%
  \BibitemOpen
  \bibfield  {author} {\bibinfo {author} {\bibfnamefont {Jacopo}\ \bibnamefont
  {Grilli}}, \bibinfo {author} {\bibfnamefont {Matteo}\ \bibnamefont
  {Adorisio}}, \bibinfo {author} {\bibfnamefont {Samir}\ \bibnamefont
  {Suweis}}, \bibinfo {author} {\bibfnamefont {György}\ \bibnamefont
  {Barabás}}, \bibinfo {author} {\bibfnamefont {Jayanth~R.}\ \bibnamefont
  {Banavar}}, \bibinfo {author} {\bibfnamefont {Stefano}\ \bibnamefont
  {Allesina}}, \ and\ \bibinfo {author} {\bibfnamefont {Amos}\ \bibnamefont
  {Maritan}},\ }\bibfield  {title} {\enquote {\bibinfo {title} {Feasibility and
  coexistence of large ecological communities},}\ }\href {\doibase
  10.1038/ncomms14389} {\bibfield  {journal} {\bibinfo  {journal} {Nature
  Communications}\ }\textbf {\bibinfo {volume} {8}},\ \bibinfo {pages} {14389}
  (\bibinfo {year} {2017})}\BibitemShut {NoStop}%
\bibitem [{\citenamefont {Mott}(1987)}]{Mott1987Mobility}%
  \BibitemOpen
  \bibfield  {author} {\bibinfo {author} {\bibfnamefont {N}~\bibnamefont
  {Mott}},\ }\bibfield  {title} {\enquote {\bibinfo {title} {The mobility edge
  since 1967},}\ }\href {\doibase 10.1088/0022-3719/20/21/008} {\bibfield
  {journal} {\bibinfo  {journal} {Journal of Physics C: Solid State Physics}\
  }\textbf {\bibinfo {volume} {20}},\ \bibinfo {pages} {3075} (\bibinfo {year}
  {1987})}\BibitemShut {NoStop}%
\bibitem [{\citenamefont {Semeghini}\ \emph {et~al.}(2015)\citenamefont
  {Semeghini}, \citenamefont {Landini}, \citenamefont {Castilho}, \citenamefont
  {Roy}, \citenamefont {Spagnolli}, \citenamefont {Trenkwalder}, \citenamefont
  {Fattori}, \citenamefont {Inguscio},\ and\ \citenamefont
  {Modugno}}]{Semeghini2015Measurement}%
  \BibitemOpen
  \bibfield  {author} {\bibinfo {author} {\bibfnamefont {G.}~\bibnamefont
  {Semeghini}}, \bibinfo {author} {\bibfnamefont {M.}~\bibnamefont {Landini}},
  \bibinfo {author} {\bibfnamefont {P.}~\bibnamefont {Castilho}}, \bibinfo
  {author} {\bibfnamefont {S.}~\bibnamefont {Roy}}, \bibinfo {author}
  {\bibfnamefont {G.}~\bibnamefont {Spagnolli}}, \bibinfo {author}
  {\bibfnamefont {A.}~\bibnamefont {Trenkwalder}}, \bibinfo {author}
  {\bibfnamefont {M.}~\bibnamefont {Fattori}}, \bibinfo {author} {\bibfnamefont
  {M.}~\bibnamefont {Inguscio}}, \ and\ \bibinfo {author} {\bibfnamefont
  {G.}~\bibnamefont {Modugno}},\ }\bibfield  {title} {\enquote {\bibinfo
  {title} {Measurement of the mobility edge for {3D} {Anderson}
  localization},}\ }\href {\doibase 10.1038/nphys3339} {\bibfield  {journal}
  {\bibinfo  {journal} {Nat. Phys}\ }\textbf {\bibinfo {volume} {11}},\
  \bibinfo {pages} {554} (\bibinfo {year} {2015})}\BibitemShut {NoStop}%
\bibitem [{\citenamefont {Pixley}\ \emph {et~al.}(2015)\citenamefont {Pixley},
  \citenamefont {Goswami},\ and\ \citenamefont
  {Das~Sarma}}]{Pixley2015Anderson}%
  \BibitemOpen
  \bibfield  {author} {\bibinfo {author} {\bibfnamefont {J.~H.}\ \bibnamefont
  {Pixley}}, \bibinfo {author} {\bibfnamefont {Pallab}\ \bibnamefont
  {Goswami}}, \ and\ \bibinfo {author} {\bibfnamefont {S.}~\bibnamefont
  {Das~Sarma}},\ }\bibfield  {title} {\enquote {\bibinfo {title} {Anderson
  localization and the quantum phase diagram of three dimensional disordered
  dirac semimetals},}\ }\href {\doibase 10.1103/PhysRevLett.115.076601}
  {\bibfield  {journal} {\bibinfo  {journal} {Phys. Rev. Lett.}\ }\textbf
  {\bibinfo {volume} {115}},\ \bibinfo {pages} {076601} (\bibinfo {year}
  {2015})}\BibitemShut {NoStop}%
\bibitem [{\citenamefont {Biddle}\ \emph {et~al.}(2009)\citenamefont {Biddle},
  \citenamefont {Wang}, \citenamefont {Priour},\ and\ \citenamefont
  {Das~Sarma}}]{Biddle2009Localization}%
  \BibitemOpen
  \bibfield  {author} {\bibinfo {author} {\bibfnamefont {J.}~\bibnamefont
  {Biddle}}, \bibinfo {author} {\bibfnamefont {B.}~\bibnamefont {Wang}},
  \bibinfo {author} {\bibfnamefont {D.~J.}\ \bibnamefont {Priour}}, \ and\
  \bibinfo {author} {\bibfnamefont {S.}~\bibnamefont {Das~Sarma}},\ }\bibfield
  {title} {\enquote {\bibinfo {title} {Localization in one-dimensional
  incommensurate lattices beyond the {Aubry}-{Andr}{\textbackslash}'e model},}\
  }\href {\doibase 10.1103/PhysRevA.80.021603} {\bibfield  {journal} {\bibinfo
  {journal} {Phys. Rev. A}\ }\textbf {\bibinfo {volume} {80}},\ \bibinfo
  {pages} {021603} (\bibinfo {year} {2009})}\BibitemShut {NoStop}%
\bibitem [{\citenamefont {Biddle}\ and\ \citenamefont
  {Das~Sarma}(2010)}]{Biddle2010Predicted}%
  \BibitemOpen
  \bibfield  {author} {\bibinfo {author} {\bibfnamefont {J.}~\bibnamefont
  {Biddle}}\ and\ \bibinfo {author} {\bibfnamefont {S.}~\bibnamefont
  {Das~Sarma}},\ }\bibfield  {title} {\enquote {\bibinfo {title} {Predicted
  {Mobility} {Edges} in {One}-{Dimensional} {Incommensurate} {Optical}
  {Lattices}: {An} {Exactly} {Solvable} {Model} of {Anderson}
  {Localization}},}\ }\href {\doibase 10.1103/PhysRevLett.104.070601}
  {\bibfield  {journal} {\bibinfo  {journal} {Phys. Rev. Lett.}\ }\textbf
  {\bibinfo {volume} {104}},\ \bibinfo {pages} {070601} (\bibinfo {year}
  {2010})}\BibitemShut {NoStop}%
\bibitem [{\citenamefont {Ganeshan}\ \emph {et~al.}(2015)\citenamefont
  {Ganeshan}, \citenamefont {Pixley},\ and\ \citenamefont
  {Das~Sarma}}]{Ganeshan2015Nearest}%
  \BibitemOpen
  \bibfield  {author} {\bibinfo {author} {\bibfnamefont {Sriram}\ \bibnamefont
  {Ganeshan}}, \bibinfo {author} {\bibfnamefont {J.~H.}\ \bibnamefont
  {Pixley}}, \ and\ \bibinfo {author} {\bibfnamefont {S.}~\bibnamefont
  {Das~Sarma}},\ }\bibfield  {title} {\enquote {\bibinfo {title} {Nearest
  {Neighbor} {Tight} {Binding} {Models} with an {Exact} {Mobility} {Edge} in
  {One} {Dimension}},}\ }\href {\doibase 10.1103/PhysRevLett.114.146601}
  {\bibfield  {journal} {\bibinfo  {journal} {Phys. Rev. Lett.}\ }\textbf
  {\bibinfo {volume} {114}},\ \bibinfo {pages} {146601} (\bibinfo {year}
  {2015})}\BibitemShut {NoStop}%
\bibitem [{\citenamefont {Wang}\ \emph {et~al.}(2020)\citenamefont {Wang},
  \citenamefont {Xia}, \citenamefont {Zhang}, \citenamefont {Yao},
  \citenamefont {Chen}, \citenamefont {You}, \citenamefont {Zhou},\ and\
  \citenamefont {Liu}}]{Wang2020one-dimensional}%
  \BibitemOpen
  \bibfield  {author} {\bibinfo {author} {\bibfnamefont {Yucheng}\ \bibnamefont
  {Wang}}, \bibinfo {author} {\bibfnamefont {Xu}~\bibnamefont {Xia}}, \bibinfo
  {author} {\bibfnamefont {Long}\ \bibnamefont {Zhang}}, \bibinfo {author}
  {\bibfnamefont {Hepeng}\ \bibnamefont {Yao}}, \bibinfo {author}
  {\bibfnamefont {Shu}\ \bibnamefont {Chen}}, \bibinfo {author} {\bibfnamefont
  {Jiangong}\ \bibnamefont {You}}, \bibinfo {author} {\bibfnamefont
  {Qi}~\bibnamefont {Zhou}}, \ and\ \bibinfo {author} {\bibfnamefont
  {Xiong-Jun}\ \bibnamefont {Liu}},\ }\bibfield  {title} {\enquote {\bibinfo
  {title} {One-{Dimensional} {Quasiperiodic} {Mosaic} {Lattice} with {Exact}
  {Mobility} {Edges}},}\ }\href {\doibase 10.1103/PhysRevLett.125.196604}
  {\bibfield  {journal} {\bibinfo  {journal} {Phys. Rev. Lett.}\ }\textbf
  {\bibinfo {volume} {125}},\ \bibinfo {pages} {196604} (\bibinfo {year}
  {2020})}\BibitemShut {NoStop}%
\bibitem [{\citenamefont {Wang}\ \emph {et~al.}(2022)\citenamefont {Wang},
  \citenamefont {Zhang}, \citenamefont {Li}, \citenamefont {Wu}, \citenamefont
  {Liu}, \citenamefont {Mei}, \citenamefont {Hu}, \citenamefont {Xiao},
  \citenamefont {Ma}, \citenamefont {Chin},\ and\ \citenamefont
  {Jia}}]{WangObservation2022}%
  \BibitemOpen
  \bibfield  {author} {\bibinfo {author} {\bibfnamefont {Yunfei}\ \bibnamefont
  {Wang}}, \bibinfo {author} {\bibfnamefont {Jia-Hui}\ \bibnamefont {Zhang}},
  \bibinfo {author} {\bibfnamefont {Yuqing}\ \bibnamefont {Li}}, \bibinfo
  {author} {\bibfnamefont {Jizhou}\ \bibnamefont {Wu}}, \bibinfo {author}
  {\bibfnamefont {Wenliang}\ \bibnamefont {Liu}}, \bibinfo {author}
  {\bibfnamefont {Feng}\ \bibnamefont {Mei}}, \bibinfo {author} {\bibfnamefont
  {Ying}\ \bibnamefont {Hu}}, \bibinfo {author} {\bibfnamefont {Liantuan}\
  \bibnamefont {Xiao}}, \bibinfo {author} {\bibfnamefont {Jie}\ \bibnamefont
  {Ma}}, \bibinfo {author} {\bibfnamefont {Cheng}\ \bibnamefont {Chin}}, \ and\
  \bibinfo {author} {\bibfnamefont {Suotang}\ \bibnamefont {Jia}},\ }\bibfield
  {title} {\enquote {\bibinfo {title} {Observation of interaction-induced
  mobility edge in an atomic aubry-andr\'e wire},}\ }\href {\doibase
  10.1103/PhysRevLett.129.103401} {\bibfield  {journal} {\bibinfo  {journal}
  {Phys. Rev. Lett.}\ }\textbf {\bibinfo {volume} {129}},\ \bibinfo {pages}
  {103401} (\bibinfo {year} {2022})}\BibitemShut {NoStop}%
\bibitem [{\citenamefont {An}\ \emph {et~al.}(2021)\citenamefont {An},
  \citenamefont {Padavi\ifmmode~\acute{c}\else \'{c}\fi{}}, \citenamefont
  {Meier}, \citenamefont {Hegde}, \citenamefont {Ganeshan}, \citenamefont
  {Pixley}, \citenamefont {Vishveshwara},\ and\ \citenamefont
  {Gadway}}]{Alex2021Interaction}%
  \BibitemOpen
  \bibfield  {author} {\bibinfo {author} {\bibfnamefont {Fangzhao~Alex}\
  \bibnamefont {An}}, \bibinfo {author} {\bibfnamefont {Karmela}\ \bibnamefont
  {Padavi\ifmmode~\acute{c}\else \'{c}\fi{}}}, \bibinfo {author} {\bibfnamefont
  {Eric~J.}\ \bibnamefont {Meier}}, \bibinfo {author} {\bibfnamefont {Suraj}\
  \bibnamefont {Hegde}}, \bibinfo {author} {\bibfnamefont {Sriram}\
  \bibnamefont {Ganeshan}}, \bibinfo {author} {\bibfnamefont {J.~H.}\
  \bibnamefont {Pixley}}, \bibinfo {author} {\bibfnamefont {Smitha}\
  \bibnamefont {Vishveshwara}}, \ and\ \bibinfo {author} {\bibfnamefont
  {Bryce}\ \bibnamefont {Gadway}},\ }\bibfield  {title} {\enquote {\bibinfo
  {title} {Interactions and mobility edges: Observing the generalized
  aubry-andr\'e model},}\ }\href {\doibase 10.1103/PhysRevLett.126.040603}
  {\bibfield  {journal} {\bibinfo  {journal} {Phys. Rev. Lett.}\ }\textbf
  {\bibinfo {volume} {126}},\ \bibinfo {pages} {040603} (\bibinfo {year}
  {2021})}\BibitemShut {NoStop}%
\bibitem [{\citenamefont {Wei}\ \emph {et~al.}(2019)\citenamefont {Wei},
  \citenamefont {Cheng}, \citenamefont {Xianlong},\ and\ \citenamefont
  {Mondaini}}]{Wei2019Investigating}%
  \BibitemOpen
  \bibfield  {author} {\bibinfo {author} {\bibfnamefont {Xingbo}\ \bibnamefont
  {Wei}}, \bibinfo {author} {\bibfnamefont {Chen}\ \bibnamefont {Cheng}},
  \bibinfo {author} {\bibfnamefont {Gao}\ \bibnamefont {Xianlong}}, \ and\
  \bibinfo {author} {\bibfnamefont {Rubem}\ \bibnamefont {Mondaini}},\
  }\bibfield  {title} {\enquote {\bibinfo {title} {Investigating many-body
  mobility edges in isolated quantum systems},}\ }\href {\doibase
  10.1103/PhysRevB.99.165137} {\bibfield  {journal} {\bibinfo  {journal} {Phys.
  Rev. B}\ }\textbf {\bibinfo {volume} {99}},\ \bibinfo {pages} {165137}
  (\bibinfo {year} {2019})}\BibitemShut {NoStop}%
\bibitem [{\citenamefont {Luitz}\ \emph {et~al.}(2015)\citenamefont {Luitz},
  \citenamefont {Laflorencie},\ and\ \citenamefont
  {Alet}}]{Luitz2015Many-body}%
  \BibitemOpen
  \bibfield  {author} {\bibinfo {author} {\bibfnamefont {David~J.}\
  \bibnamefont {Luitz}}, \bibinfo {author} {\bibfnamefont {Nicolas}\
  \bibnamefont {Laflorencie}}, \ and\ \bibinfo {author} {\bibfnamefont
  {Fabien}\ \bibnamefont {Alet}},\ }\bibfield  {title} {\enquote {\bibinfo
  {title} {Many-body localization edge in the random-field {Heisenberg}
  chain},}\ }\href {\doibase 10.1103/PhysRevB.91.081103} {\bibfield  {journal}
  {\bibinfo  {journal} {Phys. Rev. B}\ }\textbf {\bibinfo {volume} {91}},\
  \bibinfo {pages} {081103} (\bibinfo {year} {2015})}\BibitemShut {NoStop}%
\bibitem [{\citenamefont {Brighi}\ \emph {et~al.}(2020)\citenamefont {Brighi},
  \citenamefont {Abanin},\ and\ \citenamefont {Serbyn}}]{Brighi2020Stability}%
  \BibitemOpen
  \bibfield  {author} {\bibinfo {author} {\bibfnamefont {Pietro}\ \bibnamefont
  {Brighi}}, \bibinfo {author} {\bibfnamefont {Dmitry~A.}\ \bibnamefont
  {Abanin}}, \ and\ \bibinfo {author} {\bibfnamefont {Maksym}\ \bibnamefont
  {Serbyn}},\ }\bibfield  {title} {\enquote {\bibinfo {title} {Stability of
  mobility edges in disordered interacting systems},}\ }\href {\doibase
  10.1103/PhysRevB.102.060202} {\bibfield  {journal} {\bibinfo  {journal}
  {Phys. Rev. B}\ }\textbf {\bibinfo {volume} {102}},\ \bibinfo {pages}
  {060202} (\bibinfo {year} {2020})}\BibitemShut {NoStop}%
\bibitem [{\citenamefont {Deng}\ \emph {et~al.}(2017)\citenamefont {Deng},
  \citenamefont {Ganeshan}, \citenamefont {Li}, \citenamefont {Modak},
  \citenamefont {Mukerjee},\ and\ \citenamefont {Pixley}}]{Deng2017Many-body}%
  \BibitemOpen
  \bibfield  {author} {\bibinfo {author} {\bibfnamefont {Dong-Ling}\
  \bibnamefont {Deng}}, \bibinfo {author} {\bibfnamefont {Sriram}\ \bibnamefont
  {Ganeshan}}, \bibinfo {author} {\bibfnamefont {Xiaopeng}\ \bibnamefont {Li}},
  \bibinfo {author} {\bibfnamefont {Ranjan}\ \bibnamefont {Modak}}, \bibinfo
  {author} {\bibfnamefont {Subroto}\ \bibnamefont {Mukerjee}}, \ and\ \bibinfo
  {author} {\bibfnamefont {J.~H.}\ \bibnamefont {Pixley}},\ }\bibfield  {title}
  {\enquote {\bibinfo {title} {Many-body localization in incommensurate models
  with a mobility edge},}\ }\href {\doibase 10.1002/andp.201600399} {\bibfield
  {journal} {\bibinfo  {journal} {Annalen der Physik}\ }\textbf {\bibinfo
  {volume} {529}},\ \bibinfo {pages} {1600399} (\bibinfo {year}
  {2017})}\BibitemShut {NoStop}%
\bibitem [{\citenamefont {Chanda}\ \emph {et~al.}(2020)\citenamefont {Chanda},
  \citenamefont {Sierant},\ and\ \citenamefont
  {Zakrzewski}}]{Chanda2020Many-body}%
  \BibitemOpen
  \bibfield  {author} {\bibinfo {author} {\bibfnamefont {Titas}\ \bibnamefont
  {Chanda}}, \bibinfo {author} {\bibfnamefont {Piotr}\ \bibnamefont {Sierant}},
  \ and\ \bibinfo {author} {\bibfnamefont {Jakub}\ \bibnamefont {Zakrzewski}},\
  }\bibfield  {title} {\enquote {\bibinfo {title} {Many-body localization
  transition in large quantum spin chains: {The} mobility edge},}\ }\href
  {\doibase 10.1103/PhysRevResearch.2.032045} {\bibfield  {journal} {\bibinfo
  {journal} {Phys. Rev. Research}\ }\textbf {\bibinfo {volume} {2}},\ \bibinfo
  {pages} {032045} (\bibinfo {year} {2020})}\BibitemShut {NoStop}%
\bibitem [{\citenamefont {Nag}\ and\ \citenamefont
  {Garg}(2017)}]{Nag2017many-body}%
  \BibitemOpen
  \bibfield  {author} {\bibinfo {author} {\bibfnamefont {Sabyasachi}\
  \bibnamefont {Nag}}\ and\ \bibinfo {author} {\bibfnamefont {Arti}\
  \bibnamefont {Garg}},\ }\bibfield  {title} {\enquote {\bibinfo {title}
  {Many-body mobility edges in a one-dimensional system of interacting
  fermions},}\ }\href {\doibase 10.1103/PhysRevB.96.060203} {\bibfield
  {journal} {\bibinfo  {journal} {Phys. Rev. B}\ }\textbf {\bibinfo {volume}
  {96}},\ \bibinfo {pages} {060203} (\bibinfo {year} {2017})}\BibitemShut
  {NoStop}%
\bibitem [{\citenamefont {Modak}\ and\ \citenamefont
  {Mukerjee}()}]{Modak2015Many-body}%
  \BibitemOpen
  \bibfield  {author} {\bibinfo {author} {\bibfnamefont {Ranjan}\ \bibnamefont
  {Modak}}\ and\ \bibinfo {author} {\bibfnamefont {Subroto}\ \bibnamefont
  {Mukerjee}},\ }\bibfield  {title} {\enquote {\bibinfo {title} {Many-{Body}
  {Localization} in the {Presence} of a {Single}-{Particle} {Mobility}
  {Edge}},}\ }\href {\doibase 10.1103/PhysRevLett.115.230401} {\bibfield
  {journal} {\bibinfo  {journal} {Phys. Rev. Lett.}\ }\textbf {\bibinfo
  {volume} {115}},\ \bibinfo {pages} {230401}}\BibitemShut {NoStop}%
\bibitem [{\citenamefont {De~Roeck}\ \emph {et~al.}(2016)\citenamefont
  {De~Roeck}, \citenamefont {Huveneers}, \citenamefont {Müller},\ and\
  \citenamefont {Schiulaz}}]{Deroeck2016absence}%
  \BibitemOpen
  \bibfield  {author} {\bibinfo {author} {\bibfnamefont {Wojciech}\
  \bibnamefont {De~Roeck}}, \bibinfo {author} {\bibfnamefont {Francois}\
  \bibnamefont {Huveneers}}, \bibinfo {author} {\bibfnamefont {Markus}\
  \bibnamefont {Müller}}, \ and\ \bibinfo {author} {\bibfnamefont {Mauro}\
  \bibnamefont {Schiulaz}},\ }\bibfield  {title} {\enquote {\bibinfo {title}
  {Absence of many-body mobility edges},}\ }\href {\doibase
  10.1103/PhysRevB.93.014203} {\bibfield  {journal} {\bibinfo  {journal} {Phys.
  Rev. B}\ }\textbf {\bibinfo {volume} {93}},\ \bibinfo {pages} {014203}
  (\bibinfo {year} {2016})}\BibitemShut {NoStop}%
\bibitem [{\citenamefont {De~Roeck}\ and\ \citenamefont
  {Huveneers}(2017)}]{Deroeck2017Stability}%
  \BibitemOpen
  \bibfield  {author} {\bibinfo {author} {\bibfnamefont {Wojciech}\
  \bibnamefont {De~Roeck}}\ and\ \bibinfo {author} {\bibfnamefont {François}\
  \bibnamefont {Huveneers}},\ }\bibfield  {title} {\enquote {\bibinfo {title}
  {Stability and instability towards delocalization in {MBL} systems},}\ }\href
  {\doibase 10.1103/PhysRevB.95.155129} {\bibfield  {journal} {\bibinfo
  {journal} {Phys. Rev. B}\ }\textbf {\bibinfo {volume} {95}},\ \bibinfo
  {pages} {155129} (\bibinfo {year} {2017})}\BibitemShut {NoStop}%
\bibitem [{\citenamefont {Zhang}\ \emph {et~al.}(2022)\citenamefont {Zhang},
  \citenamefont {Zhou}, \citenamefont {Hu},\ and\ \citenamefont
  {Chen}}]{Zhang2022Localization}%
  \BibitemOpen
  \bibfield  {author} {\bibinfo {author} {\bibfnamefont {Yu}~\bibnamefont
  {Zhang}}, \bibinfo {author} {\bibfnamefont {Bozhen}\ \bibnamefont {Zhou}},
  \bibinfo {author} {\bibfnamefont {Haiping}\ \bibnamefont {Hu}}, \ and\
  \bibinfo {author} {\bibfnamefont {Shu}\ \bibnamefont {Chen}},\ }\bibfield
  {title} {\enquote {\bibinfo {title} {Localization, multifractality, and
  many-body localization in periodically kicked quasiperiodic lattices},}\
  }\href {\doibase 10.1103/PhysRevB.106.054312} {\bibfield  {journal} {\bibinfo
   {journal} {Phys. Rev. B}\ }\textbf {\bibinfo {volume} {106}},\ \bibinfo
  {pages} {054312} (\bibinfo {year} {2022})}\BibitemShut {NoStop}%
\bibitem [{\citenamefont {Lazarides}\ \emph {et~al.}(2015)\citenamefont
  {Lazarides}, \citenamefont {Das},\ and\ \citenamefont
  {Moessner}}]{Lazarides2015Fate}%
  \BibitemOpen
  \bibfield  {author} {\bibinfo {author} {\bibfnamefont {Achilleas}\
  \bibnamefont {Lazarides}}, \bibinfo {author} {\bibfnamefont {Arnab}\
  \bibnamefont {Das}}, \ and\ \bibinfo {author} {\bibfnamefont {Roderich}\
  \bibnamefont {Moessner}},\ }\bibfield  {title} {\enquote {\bibinfo {title}
  {Fate of {Many}-{Body} {Localization} {Under} {Periodic} {Driving}},}\ }\href
  {\doibase 10.1103/PhysRevLett.115.030402} {\bibfield  {journal} {\bibinfo
  {journal} {Phys. Rev. Lett.}\ }\textbf {\bibinfo {volume} {115}},\ \bibinfo
  {pages} {030402} (\bibinfo {year} {2015})}\BibitemShut {NoStop}%
\bibitem [{\citenamefont {Lüschen}\ \emph {et~al.}(2018)\citenamefont
  {Lüschen}, \citenamefont {Scherg}, \citenamefont {Kohlert}, \citenamefont
  {Schreiber}, \citenamefont {Bordia}, \citenamefont {Li}, \citenamefont
  {Das~Sarma},\ and\ \citenamefont {Bloch}}]{Luschen2018Single-particle}%
  \BibitemOpen
  \bibfield  {author} {\bibinfo {author} {\bibfnamefont {Henrik~P.}\
  \bibnamefont {Lüschen}}, \bibinfo {author} {\bibfnamefont {Sebastian}\
  \bibnamefont {Scherg}}, \bibinfo {author} {\bibfnamefont {Thomas}\
  \bibnamefont {Kohlert}}, \bibinfo {author} {\bibfnamefont {Michael}\
  \bibnamefont {Schreiber}}, \bibinfo {author} {\bibfnamefont {Pranjal}\
  \bibnamefont {Bordia}}, \bibinfo {author} {\bibfnamefont {Xiao}\ \bibnamefont
  {Li}}, \bibinfo {author} {\bibfnamefont {S.}~\bibnamefont {Das~Sarma}}, \
  and\ \bibinfo {author} {\bibfnamefont {Immanuel}\ \bibnamefont {Bloch}},\
  }\bibfield  {title} {\enquote {\bibinfo {title} {Single-{Particle} {Mobility}
  {Edge} in a {One}-{Dimensional} {Quasiperiodic} {Optical} {Lattice}},}\
  }\href {\doibase 10.1103/PhysRevLett.120.160404} {\bibfield  {journal}
  {\bibinfo  {journal} {Phys. Rev. Lett.}\ }\textbf {\bibinfo {volume} {120}},\
  \bibinfo {pages} {160404} (\bibinfo {year} {2018})}\BibitemShut {NoStop}%
\bibitem [{\citenamefont {Kohlert}\ \emph {et~al.}(2019)\citenamefont
  {Kohlert}, \citenamefont {Scherg}, \citenamefont {Li}, \citenamefont
  {Lüschen}, \citenamefont {Das~Sarma}, \citenamefont {Bloch},\ and\
  \citenamefont {Aidelsburger}}]{Kohlert2019Observation}%
  \BibitemOpen
  \bibfield  {author} {\bibinfo {author} {\bibfnamefont {Thomas}\ \bibnamefont
  {Kohlert}}, \bibinfo {author} {\bibfnamefont {Sebastian}\ \bibnamefont
  {Scherg}}, \bibinfo {author} {\bibfnamefont {Xiao}\ \bibnamefont {Li}},
  \bibinfo {author} {\bibfnamefont {Henrik~P.}\ \bibnamefont {Lüschen}},
  \bibinfo {author} {\bibfnamefont {Sankar}\ \bibnamefont {Das~Sarma}},
  \bibinfo {author} {\bibfnamefont {Immanuel}\ \bibnamefont {Bloch}}, \ and\
  \bibinfo {author} {\bibfnamefont {Monika}\ \bibnamefont {Aidelsburger}},\
  }\bibfield  {title} {\enquote {\bibinfo {title} {Observation of {Many}-{Body}
  {Localization} in a {One}-{Dimensional} {System} with a {Single}-{Particle}
  {Mobility} {Edge}},}\ }\href {\doibase 10.1103/PhysRevLett.122.170403}
  {\bibfield  {journal} {\bibinfo  {journal} {Phys. Rev. Lett.}\ }\textbf
  {\bibinfo {volume} {122}},\ \bibinfo {pages} {170403} (\bibinfo {year}
  {2019})}\BibitemShut {NoStop}%
\bibitem [{\citenamefont {Cizeau}\ and\ \citenamefont
  {Bouchaud}(1994)}]{Cizeau1994Theory}%
  \BibitemOpen
  \bibfield  {author} {\bibinfo {author} {\bibfnamefont {P.}~\bibnamefont
  {Cizeau}}\ and\ \bibinfo {author} {\bibfnamefont {J.~P.}\ \bibnamefont
  {Bouchaud}},\ }\bibfield  {title} {\enquote {\bibinfo {title} {Theory of
  {Lévy} matrices},}\ }\href {\doibase 10.1103/PhysRevE.50.1810} {\bibfield
  {journal} {\bibinfo  {journal} {Phys. Rev. E}\ }\textbf {\bibinfo {volume}
  {50}},\ \bibinfo {pages} {1810} (\bibinfo {year} {1994})}\BibitemShut
  {NoStop}%
\bibitem [{\citenamefont {Tarquini}\ \emph {et~al.}(2016)\citenamefont
  {Tarquini}, \citenamefont {Biroli},\ and\ \citenamefont
  {Tarzia}}]{tarquini2016level}%
  \BibitemOpen
  \bibfield  {author} {\bibinfo {author} {\bibfnamefont {E.}~\bibnamefont
  {Tarquini}}, \bibinfo {author} {\bibfnamefont {G.}~\bibnamefont {Biroli}}, \
  and\ \bibinfo {author} {\bibfnamefont {M.}~\bibnamefont {Tarzia}},\
  }\bibfield  {title} {\enquote {\bibinfo {title} {Level {Statistics} and
  {Localization} {Transitions} of {L}{\textbackslash}'evy {Matrices}},}\ }\href
  {\doibase 10.1103/PhysRevLett.116.010601} {\bibfield  {journal} {\bibinfo
  {journal} {Phys. Rev. Lett.}\ }\textbf {\bibinfo {volume} {116}},\ \bibinfo
  {pages} {010601} (\bibinfo {year} {2016})}\BibitemShut {NoStop}%
\bibitem [{\citenamefont {Biroli}\ and\ \citenamefont
  {Tarzia}(2021)}]{Biroli2021Levy}%
  \BibitemOpen
  \bibfield  {author} {\bibinfo {author} {\bibfnamefont {G.}~\bibnamefont
  {Biroli}}\ and\ \bibinfo {author} {\bibfnamefont {M.}~\bibnamefont
  {Tarzia}},\ }\bibfield  {title} {\enquote {\bibinfo {title}
  {{L\'evy-Rosenzweig-Porter random matrix ensemble}},}\ }\href {\doibase
  10.1103/PhysRevB.103.104205} {\bibfield  {journal} {\bibinfo  {journal}
  {Phys. Rev. B}\ }\textbf {\bibinfo {volume} {103}},\ \bibinfo {pages}
  {104205} (\bibinfo {year} {2021})}\BibitemShut {NoStop}%
\bibitem [{\citenamefont {Aggarwal}\ \emph {et~al.}(2019)\citenamefont
  {Aggarwal}, \citenamefont {Lopatto},\ and\ \citenamefont
  {Yau}}]{aggarwal2019goe}%
  \BibitemOpen
  \bibfield  {author} {\bibinfo {author} {\bibfnamefont {Amol}\ \bibnamefont
  {Aggarwal}}, \bibinfo {author} {\bibfnamefont {Patrick}\ \bibnamefont
  {Lopatto}}, \ and\ \bibinfo {author} {\bibfnamefont {Horng-Tzer}\
  \bibnamefont {Yau}},\ }\href {http://arxiv.org/abs/1806.07363} {\enquote
  {\bibinfo {title} {{GOE} {Statistics} for {Levy} {Matrices}},}\ } (\bibinfo
  {year} {2019}),\ \bibinfo {note} {arXiv:1806.07363 [math-ph]}\BibitemShut
  {NoStop}%
\bibitem [{\citenamefont {Sarkar}\ \emph {et~al.}(2023)\citenamefont {Sarkar},
  \citenamefont {Ghosh},\ and\ \citenamefont {Khaymovich}}]{Sarkar2023Tuning}%
  \BibitemOpen
  \bibfield  {author} {\bibinfo {author} {\bibfnamefont {Madhumita}\
  \bibnamefont {Sarkar}}, \bibinfo {author} {\bibfnamefont {Roopayan}\
  \bibnamefont {Ghosh}}, \ and\ \bibinfo {author} {\bibfnamefont {Ivan~M.}\
  \bibnamefont {Khaymovich}},\ }\bibfield  {title} {\enquote {\bibinfo {title}
  {Tuning the phase diagram of a rosenzweig-porter model with fractal
  disorder},}\ }\href {\doibase 10.1103/PhysRevB.108.L060203} {\bibfield
  {journal} {\bibinfo  {journal} {Phys. Rev. B}\ }\textbf {\bibinfo {volume}
  {108}},\ \bibinfo {pages} {L060203} (\bibinfo {year} {2023})}\BibitemShut
  {NoStop}%
\bibitem [{SI()}]{SI}%
  \BibitemOpen
  \href@noop {} {}\bibinfo {note} {See Supplemental Material for (I)
  Characterization of the Rosenzweig-Porter (RP) ensemble (II) Phase diagram of
  the sparse Rosenzweig-Porter ensemble. (III) The derivation of Eq. 4 and Eq.
  5 in the main text. (IV) Finite-size analysis of the phase boundary. (V)
  Coupled random matrice with Gaussian unitary ensembles.}\BibitemShut {Stop}%
\bibitem [{\citenamefont {Rosenzweig}\ and\ \citenamefont
  {Porter}(1960)}]{Rosenzweig1960REpulsion}%
  \BibitemOpen
  \bibfield  {author} {\bibinfo {author} {\bibfnamefont {Norbert}\ \bibnamefont
  {Rosenzweig}}\ and\ \bibinfo {author} {\bibfnamefont {Charles~E.}\
  \bibnamefont {Porter}},\ }\bibfield  {title} {\enquote {\bibinfo {title}
  {"repulsion of energy levels" in complex atomic spectra},}\ }\href {\doibase
  10.1103/PhysRev.120.1698} {\bibfield  {journal} {\bibinfo  {journal} {Phys.
  Rev.}\ }\textbf {\bibinfo {volume} {120}},\ \bibinfo {pages} {1698--1714}
  (\bibinfo {year} {1960})}\BibitemShut {NoStop}%
\bibitem [{\citenamefont {Kravtsov}\ \emph {et~al.}(2015)\citenamefont
  {Kravtsov}, \citenamefont {Khaymovich}, \citenamefont {Cuevas},\ and\
  \citenamefont {Amini}}]{Kravtsov2015Random}%
  \BibitemOpen
  \bibfield  {author} {\bibinfo {author} {\bibfnamefont {V~E}\ \bibnamefont
  {Kravtsov}}, \bibinfo {author} {\bibfnamefont {I~M}\ \bibnamefont
  {Khaymovich}}, \bibinfo {author} {\bibfnamefont {E}~\bibnamefont {Cuevas}}, \
  and\ \bibinfo {author} {\bibfnamefont {M}~\bibnamefont {Amini}},\ }\bibfield
  {title} {\enquote {\bibinfo {title} {A random matrix model with localization
  and ergodic transitions},}\ }\href {\doibase 10.1088/1367-2630/17/12/122002}
  {\bibfield  {journal} {\bibinfo  {journal} {New J. Phys.}\ }\textbf {\bibinfo
  {volume} {17}},\ \bibinfo {pages} {122002} (\bibinfo {year}
  {2015})}\BibitemShut {NoStop}%
\bibitem [{\citenamefont {Bogomolny}\ and\ \citenamefont
  {Sieber}(2018)}]{Bogomolny2018Eigenfunction}%
  \BibitemOpen
  \bibfield  {author} {\bibinfo {author} {\bibfnamefont {E.}~\bibnamefont
  {Bogomolny}}\ and\ \bibinfo {author} {\bibfnamefont {M.}~\bibnamefont
  {Sieber}},\ }\bibfield  {title} {\enquote {\bibinfo {title} {{Eigenfunction
  distribution for the {Rosenzweig}-{Porter} model}},}\ }\href {\doibase
  10.1103/PhysRevE.98.032139} {\bibfield  {journal} {\bibinfo  {journal} {Phys.
  Rev. E}\ }\textbf {\bibinfo {volume} {98}},\ \bibinfo {pages} {032139}
  (\bibinfo {year} {2018})}\BibitemShut {NoStop}%
\bibitem [{\citenamefont {Facoetti}\ \emph {et~al.}(2016)\citenamefont
  {Facoetti}, \citenamefont {Vivo},\ and\ \citenamefont
  {Biroli}}]{Facoetti2016Non-ergodic}%
  \BibitemOpen
  \bibfield  {author} {\bibinfo {author} {\bibfnamefont {Davide}\ \bibnamefont
  {Facoetti}}, \bibinfo {author} {\bibfnamefont {Pierpaolo}\ \bibnamefont
  {Vivo}}, \ and\ \bibinfo {author} {\bibfnamefont {Giulio}\ \bibnamefont
  {Biroli}},\ }\bibfield  {title} {\enquote {\bibinfo {title} {From non-ergodic
  eigenvectors to local resolvent statistics and back: {A} random matrix
  perspective},}\ }\href {\doibase 10.1209/0295-5075/115/47003} {\bibfield
  {journal} {\bibinfo  {journal} {EPL}\ }\textbf {\bibinfo {volume} {115}},\
  \bibinfo {pages} {47003} (\bibinfo {year} {2016})}\BibitemShut {NoStop}%
\bibitem [{\citenamefont {Khaymovich}\ \emph {et~al.}(2020)\citenamefont
  {Khaymovich}, \citenamefont {Kravtsov}, \citenamefont {Altshuler},\ and\
  \citenamefont {Ioffe}}]{Khaymovich2020Fragile}%
  \BibitemOpen
  \bibfield  {author} {\bibinfo {author} {\bibfnamefont {I.~M.}\ \bibnamefont
  {Khaymovich}}, \bibinfo {author} {\bibfnamefont {V.~E.}\ \bibnamefont
  {Kravtsov}}, \bibinfo {author} {\bibfnamefont {B.~L.}\ \bibnamefont
  {Altshuler}}, \ and\ \bibinfo {author} {\bibfnamefont {L.~B.}\ \bibnamefont
  {Ioffe}},\ }\bibfield  {title} {\enquote {\bibinfo {title} {{Fragile extended
  phases in the log-normal Rosenzweig-Porter model}},}\ }\href {\doibase
  10.1103/PhysRevResearch.2.043346} {\bibfield  {journal} {\bibinfo  {journal}
  {Phys. Rev. Res.}\ }\textbf {\bibinfo {volume} {2}},\ \bibinfo {pages}
  {043346} (\bibinfo {year} {2020})}\BibitemShut {NoStop}%
\bibitem [{\citenamefont {De~Tomasi}\ and\ \citenamefont
  {Khaymovich}(2022)}]{Tomasi2022Non-Hermitian}%
  \BibitemOpen
  \bibfield  {author} {\bibinfo {author} {\bibfnamefont {Giuseppe}\
  \bibnamefont {De~Tomasi}}\ and\ \bibinfo {author} {\bibfnamefont {Ivan~M.}\
  \bibnamefont {Khaymovich}},\ }\bibfield  {title} {\enquote {\bibinfo {title}
  {{Non-Hermitian Rosenzweig-Porter random-matrix ensemble: Obstruction to the
  fractal phase}},}\ }\href {\doibase 10.1103/PhysRevB.106.094204} {\bibfield
  {journal} {\bibinfo  {journal} {Phys. Rev. B}\ }\textbf {\bibinfo {volume}
  {106}},\ \bibinfo {pages} {094204} (\bibinfo {year} {2022})}\BibitemShut
  {NoStop}%
\bibitem [{\citenamefont {Buijsman}\ and\ \citenamefont
  {Lev}(2022)}]{Wouter2022Circular}%
  \BibitemOpen
  \bibfield  {author} {\bibinfo {author} {\bibfnamefont {Wouter}\ \bibnamefont
  {Buijsman}}\ and\ \bibinfo {author} {\bibfnamefont {Yevgeny~Bar}\
  \bibnamefont {Lev}},\ }\bibfield  {title} {\enquote {\bibinfo {title}
  {{Circular Rosenzweig-Porter random matrix ensemble}},}\ }\href {\doibase
  10.21468/SciPostPhys.12.3.082} {\bibfield  {journal} {\bibinfo  {journal}
  {SciPost Phys.}\ }\textbf {\bibinfo {volume} {12}},\ \bibinfo {pages} {082}
  (\bibinfo {year} {2022})}\BibitemShut {NoStop}%
\bibitem [{\citenamefont {Das}\ and\ \citenamefont
  {Ghosh}(2022)}]{Das2022Nonergodic}%
  \BibitemOpen
  \bibfield  {author} {\bibinfo {author} {\bibfnamefont {Adway~Kumar}\
  \bibnamefont {Das}}\ and\ \bibinfo {author} {\bibfnamefont {Anandamohan}\
  \bibnamefont {Ghosh}},\ }\bibfield  {title} {\enquote {\bibinfo {title}
  {Nonergodic extended states in the $\ensuremath{\beta}$ ensemble},}\ }\href
  {\doibase 10.1103/PhysRevE.105.054121} {\bibfield  {journal} {\bibinfo
  {journal} {Phys. Rev. E}\ }\textbf {\bibinfo {volume} {105}},\ \bibinfo
  {pages} {054121} (\bibinfo {year} {2022})}\BibitemShut {NoStop}%
\bibitem [{\citenamefont {Das}\ \emph {et~al.}(2023)\citenamefont {Das},
  \citenamefont {Ghosh},\ and\ \citenamefont {Khaymovich}}]{das2023absence}%
  \BibitemOpen
  \bibfield  {author} {\bibinfo {author} {\bibfnamefont {Adway~Kumar}\
  \bibnamefont {Das}}, \bibinfo {author} {\bibfnamefont {Anandamohan}\
  \bibnamefont {Ghosh}}, \ and\ \bibinfo {author} {\bibfnamefont {Ivan~M.}\
  \bibnamefont {Khaymovich}},\ }\href@noop {} {\enquote {\bibinfo {title}
  {Absence of mobility edge in short-range uncorrelated disordered model:
  Coexistence of localized and extended states},}\ } (\bibinfo {year} {2023}),\
  \Eprint {http://arxiv.org/abs/2305.02351} {arXiv:2305.02351
  [cond-mat.dis-nn]} \BibitemShut {NoStop}%
\bibitem [{\citenamefont {Giraud}\ \emph {et~al.}(2022)\citenamefont {Giraud},
  \citenamefont {Mac\'e}, \citenamefont {Vernier},\ and\ \citenamefont
  {Alet}}]{Giraud2022Probing}%
  \BibitemOpen
  \bibfield  {author} {\bibinfo {author} {\bibfnamefont {Olivier}\ \bibnamefont
  {Giraud}}, \bibinfo {author} {\bibfnamefont {Nicolas}\ \bibnamefont
  {Mac\'e}}, \bibinfo {author} {\bibfnamefont {\'Eric}\ \bibnamefont
  {Vernier}}, \ and\ \bibinfo {author} {\bibfnamefont {Fabien}\ \bibnamefont
  {Alet}},\ }\bibfield  {title} {\enquote {\bibinfo {title} {Probing symmetries
  of quantum many-body systems through gap ratio statistics},}\ }\href
  {\doibase 10.1103/PhysRevX.12.011006} {\bibfield  {journal} {\bibinfo
  {journal} {Phys. Rev. X}\ }\textbf {\bibinfo {volume} {12}},\ \bibinfo
  {pages} {011006} (\bibinfo {year} {2022})}\BibitemShut {NoStop}%
\bibitem [{\citenamefont {Thiery}\ \emph {et~al.}(2018)\citenamefont {Thiery},
  \citenamefont {Huveneers}, \citenamefont {M{\"u}ller},\ and\ \citenamefont
  {De~Roeck}}]{Thiery2018Many}%
  \BibitemOpen
  \bibfield  {author} {\bibinfo {author} {\bibfnamefont {Thimoth{\'e}e}\
  \bibnamefont {Thiery}}, \bibinfo {author} {\bibfnamefont {Fran{\c{c}}ois}\
  \bibnamefont {Huveneers}}, \bibinfo {author} {\bibfnamefont {Markus}\
  \bibnamefont {M{\"u}ller}}, \ and\ \bibinfo {author} {\bibfnamefont
  {Wojciech}\ \bibnamefont {De~Roeck}},\ }\bibfield  {title} {\enquote
  {\bibinfo {title} {Many-body delocalization as a quantum avalanche},}\
  }\href@noop {} {\bibfield  {journal} {\bibinfo  {journal} {Phys. Rev. Lett.}\
  }\textbf {\bibinfo {volume} {121}},\ \bibinfo {pages} {140601} (\bibinfo
  {year} {2018})}\BibitemShut {NoStop}%
\bibitem [{\citenamefont {Rubio-Abadal}\ \emph {et~al.}(2019)\citenamefont
  {Rubio-Abadal}, \citenamefont {Choi}, \citenamefont {Zeiher}, \citenamefont
  {Hollerith}, \citenamefont {Rui}, \citenamefont {Bloch},\ and\ \citenamefont
  {Gross}}]{Rubio2019Many}%
  \BibitemOpen
  \bibfield  {author} {\bibinfo {author} {\bibfnamefont {Antonio}\ \bibnamefont
  {Rubio-Abadal}}, \bibinfo {author} {\bibfnamefont {Jae-yoon}\ \bibnamefont
  {Choi}}, \bibinfo {author} {\bibfnamefont {Johannes}\ \bibnamefont {Zeiher}},
  \bibinfo {author} {\bibfnamefont {Simon}\ \bibnamefont {Hollerith}}, \bibinfo
  {author} {\bibfnamefont {Jun}\ \bibnamefont {Rui}}, \bibinfo {author}
  {\bibfnamefont {Immanuel}\ \bibnamefont {Bloch}}, \ and\ \bibinfo {author}
  {\bibfnamefont {Christian}\ \bibnamefont {Gross}},\ }\bibfield  {title}
  {\enquote {\bibinfo {title} {Many-body delocalization in the presence of a
  quantum bath},}\ }\href@noop {} {\bibfield  {journal} {\bibinfo  {journal}
  {Phys. Rev. X}\ }\textbf {\bibinfo {volume} {9}},\ \bibinfo {pages} {041014}
  (\bibinfo {year} {2019})}\BibitemShut {NoStop}%
\bibitem [{\citenamefont {See~Toh}\ \emph {et~al.}(2022)\citenamefont
  {See~Toh}, \citenamefont {McCormick}, \citenamefont {Tang}, \citenamefont
  {Su}, \citenamefont {Luo}, \citenamefont {Zhang},\ and\ \citenamefont
  {Gupta}}]{See2022Many}%
  \BibitemOpen
  \bibfield  {author} {\bibinfo {author} {\bibfnamefont {Jun~Hui}\ \bibnamefont
  {See~Toh}}, \bibinfo {author} {\bibfnamefont {Katherine~C}\ \bibnamefont
  {McCormick}}, \bibinfo {author} {\bibfnamefont {Xinxin}\ \bibnamefont
  {Tang}}, \bibinfo {author} {\bibfnamefont {Ying}\ \bibnamefont {Su}},
  \bibinfo {author} {\bibfnamefont {Xi-Wang}\ \bibnamefont {Luo}}, \bibinfo
  {author} {\bibfnamefont {Chuanwei}\ \bibnamefont {Zhang}}, \ and\ \bibinfo
  {author} {\bibfnamefont {Subhadeep}\ \bibnamefont {Gupta}},\ }\bibfield
  {title} {\enquote {\bibinfo {title} {Many-body dynamical delocalization in a
  kicked one-dimensional ultracold gas},}\ }\href@noop {} {\bibfield  {journal}
  {\bibinfo  {journal} {Nat. Phys.}\ }\textbf {\bibinfo {volume} {18}},\
  \bibinfo {pages} {1297--1301} (\bibinfo {year} {2022})}\BibitemShut {NoStop}%
\bibitem [{Not()}]{NoteCentralLimit}%
  \BibitemOpen
  \href@noop {} {}\bibinfo {note} {The other choices of random entries with
  finite variance are equivalent. This is due to the central limit theorem,
  where the sum of all identical distributed random entries converges to the
  Gaussian function. When the variance of the random entries does not exist,
  the model is equivalent to the L\'{e}vy random matrix.}\BibitemShut {Stop}%
\bibitem [{\citenamefont {Sakurai}\ and\ \citenamefont
  {Commins}(1995)}]{sakurai1995modern}%
  \BibitemOpen
  \bibfield  {author} {\bibinfo {author} {\bibfnamefont {Jun~John}\
  \bibnamefont {Sakurai}}\ and\ \bibinfo {author} {\bibfnamefont {Eugene~D}\
  \bibnamefont {Commins}},\ }\href@noop {} {\enquote {\bibinfo {title} {Modern
  quantum mechanics, revised edition},}\ } (\bibinfo {year} {1995})\BibitemShut
  {NoStop}%
\bibitem [{\citenamefont {Micklitz}\ \emph {et~al.}(2022)\citenamefont
  {Micklitz}, \citenamefont {Morningstar}, \citenamefont {Altland},\ and\
  \citenamefont {Huse}}]{Micklitz2022Emergence}%
  \BibitemOpen
  \bibfield  {author} {\bibinfo {author} {\bibfnamefont {Tobias}\ \bibnamefont
  {Micklitz}}, \bibinfo {author} {\bibfnamefont {Alan}\ \bibnamefont
  {Morningstar}}, \bibinfo {author} {\bibfnamefont {Alexander}\ \bibnamefont
  {Altland}}, \ and\ \bibinfo {author} {\bibfnamefont {David~A.}\ \bibnamefont
  {Huse}},\ }\bibfield  {title} {\enquote {\bibinfo {title} {Emergence of
  fermi's golden rule},}\ }\href {\doibase 10.1103/PhysRevLett.129.140402}
  {\bibfield  {journal} {\bibinfo  {journal} {Phys. Rev. Lett.}\ }\textbf
  {\bibinfo {volume} {129}},\ \bibinfo {pages} {140402} (\bibinfo {year}
  {2022})}\BibitemShut {NoStop}%
\bibitem [{\citenamefont {Venturelli}\ \emph {et~al.}(2023)\citenamefont
  {Venturelli}, \citenamefont {Cugliandolo}, \citenamefont {Schehr},\ and\
  \citenamefont {Tarzia}}]{Venturelli2023Replica}%
  \BibitemOpen
  \bibfield  {author} {\bibinfo {author} {\bibfnamefont {Davide}\ \bibnamefont
  {Venturelli}}, \bibinfo {author} {\bibfnamefont {Leticia~F.}\ \bibnamefont
  {Cugliandolo}}, \bibinfo {author} {\bibfnamefont {Grégory}\ \bibnamefont
  {Schehr}}, \ and\ \bibinfo {author} {\bibfnamefont {Marco}\ \bibnamefont
  {Tarzia}},\ }\bibfield  {title} {\enquote {\bibinfo {title} {{Replica
  approach to the generalized Rosenzweig-Porter model}},}\ }\href {\doibase
  10.21468/SciPostPhys.14.5.110} {\bibfield  {journal} {\bibinfo  {journal}
  {SciPost Phys.}\ }\textbf {\bibinfo {volume} {14}},\ \bibinfo {pages} {110}
  (\bibinfo {year} {2023})}\BibitemShut {NoStop}%
\bibitem [{RPn()}]{RPnoteofExponent}%
  \BibitemOpen
  \href@noop {} {}\bibinfo {note} {For a single random matrix, the criterion of
  $\langle \sum_j |H_{ij}|\rangle$ will yields $N^{\nu - \gamma/2}$.
  Furthermore, we have $(\sum_j |H_{ij}|)^2 \sim N^{\nu} \sum_j |H_{ij}|^2$.
  With these two results, we can obtain Eq. 6 in the main text.}\BibitemShut
  {Stop}%
\bibitem [{\citenamefont {Mirlin}\ and\ \citenamefont
  {Fyodorov}(1991)}]{mirlin1991universality}%
  \BibitemOpen
  \bibfield  {author} {\bibinfo {author} {\bibfnamefont {A.~D.}\ \bibnamefont
  {Mirlin}}\ and\ \bibinfo {author} {\bibfnamefont {Y.~V.}\ \bibnamefont
  {Fyodorov}},\ }\bibfield  {title} {\enquote {\bibinfo {title} {Universality
  of level correlation function of sparse random matrices},}\ }\href {\doibase
  10.1088/0305-4470/24/10/016} {\bibfield  {journal} {\bibinfo  {journal} {J.
  Phys. A: Math. Gen.}\ }\textbf {\bibinfo {volume} {24}},\ \bibinfo {pages}
  {2273} (\bibinfo {year} {1991})}\BibitemShut {NoStop}%
\bibitem [{\citenamefont {Fyodorov}\ and\ \citenamefont
  {Mirlin}(1991)}]{fyodorov1991localization}%
  \BibitemOpen
  \bibfield  {author} {\bibinfo {author} {\bibfnamefont {Yan~V.}\ \bibnamefont
  {Fyodorov}}\ and\ \bibinfo {author} {\bibfnamefont {Alexander~D.}\
  \bibnamefont {Mirlin}},\ }\bibfield  {title} {\enquote {\bibinfo {title}
  {Localization in ensemble of sparse random matrices},}\ }\href {\doibase
  10.1103/PhysRevLett.67.2049} {\bibfield  {journal} {\bibinfo  {journal}
  {Phys. Rev. Lett.}\ }\textbf {\bibinfo {volume} {67}},\ \bibinfo {pages}
  {2049} (\bibinfo {year} {1991})}\BibitemShut {NoStop}%
\bibitem [{\citenamefont {Berry}\ and\ \citenamefont
  {Robnik}(1984)}]{Berry1984Semiclassical}%
  \BibitemOpen
  \bibfield  {author} {\bibinfo {author} {\bibfnamefont {M.~V.}\ \bibnamefont
  {Berry}}\ and\ \bibinfo {author} {\bibfnamefont {M.}~\bibnamefont {Robnik}},\
  }\bibfield  {title} {\enquote {\bibinfo {title} {Semiclassical level spacings
  when regular and chaotic orbits coexist},}\ }\href {\doibase
  10.1088/0305-4470/17/12/013} {\bibfield  {journal} {\bibinfo  {journal} {J.
  Phys. A: Math. Gen.}\ }\textbf {\bibinfo {volume} {17}},\ \bibinfo {pages}
  {2413} (\bibinfo {year} {1984})}\BibitemShut {NoStop}%
\bibitem [{\citenamefont {Dyson}(1962{\natexlab{b}})}]{dyson1962brownian}%
  \BibitemOpen
  \bibfield  {author} {\bibinfo {author} {\bibfnamefont {Freeman~J}\
  \bibnamefont {Dyson}},\ }\bibfield  {title} {\enquote {\bibinfo {title} {{A
  Brownian-motion model for the eigenvalues of a random matrix}},}\ }\href
  {\doibase https://doi.org/10.1063/1.1703862} {\bibfield  {journal} {\bibinfo
  {journal} {J. Math. Phys.}\ }\textbf {\bibinfo {volume} {3}},\ \bibinfo
  {pages} {1191} (\bibinfo {year} {1962}{\natexlab{b}})}\BibitemShut {NoStop}%
\bibitem [{\citenamefont {Hamazaki}\ \emph {et~al.}(2019)\citenamefont
  {Hamazaki}, \citenamefont {Kawabata},\ and\ \citenamefont
  {Ueda}}]{Hamazaki2019Non-hermitian}%
  \BibitemOpen
  \bibfield  {author} {\bibinfo {author} {\bibfnamefont {Ryusuke}\ \bibnamefont
  {Hamazaki}}, \bibinfo {author} {\bibfnamefont {Kohei}\ \bibnamefont
  {Kawabata}}, \ and\ \bibinfo {author} {\bibfnamefont {Masahito}\ \bibnamefont
  {Ueda}},\ }\bibfield  {title} {\enquote {\bibinfo {title} {Non-hermitian
  many-body localization},}\ }\href {\doibase 10.1103/PhysRevLett.123.090603}
  {\bibfield  {journal} {\bibinfo  {journal} {Phys. Rev. Lett.}\ }\textbf
  {\bibinfo {volume} {123}},\ \bibinfo {pages} {090603} (\bibinfo {year}
  {2019})}\BibitemShut {NoStop}%
\bibitem [{\citenamefont {Gong}\ \emph {et~al.}(2018)\citenamefont {Gong},
  \citenamefont {Ashida}, \citenamefont {Kawabata}, \citenamefont {Takasan},
  \citenamefont {Higashikawa},\ and\ \citenamefont
  {Ueda}}]{Gong2018Topological}%
  \BibitemOpen
  \bibfield  {author} {\bibinfo {author} {\bibfnamefont {Zongping}\
  \bibnamefont {Gong}}, \bibinfo {author} {\bibfnamefont {Yuto}\ \bibnamefont
  {Ashida}}, \bibinfo {author} {\bibfnamefont {Kohei}\ \bibnamefont
  {Kawabata}}, \bibinfo {author} {\bibfnamefont {Kazuaki}\ \bibnamefont
  {Takasan}}, \bibinfo {author} {\bibfnamefont {Sho}\ \bibnamefont
  {Higashikawa}}, \ and\ \bibinfo {author} {\bibfnamefont {Masahito}\
  \bibnamefont {Ueda}},\ }\bibfield  {title} {\enquote {\bibinfo {title}
  {Topological phases of non-hermitian systems},}\ }\href {\doibase
  10.1103/PhysRevX.8.031079} {\bibfield  {journal} {\bibinfo  {journal} {Phys.
  Rev. X}\ }\textbf {\bibinfo {volume} {8}},\ \bibinfo {pages} {031079}
  (\bibinfo {year} {2018})}\BibitemShut {NoStop}%
\bibitem [{\citenamefont {Ashida}\ \emph {et~al.}(2021)\citenamefont {Ashida},
  \citenamefont {Gong},\ and\ \citenamefont {Ueda}}]{Ashida2021Nonermite}%
  \BibitemOpen
  \bibfield  {author} {\bibinfo {author} {\bibfnamefont {Yuto}\ \bibnamefont
  {Ashida}}, \bibinfo {author} {\bibfnamefont {Zongping}\ \bibnamefont {Gong}},
  \ and\ \bibinfo {author} {\bibfnamefont {Masahito}\ \bibnamefont {Ueda}},\
  }\bibfield  {title} {\enquote {\bibinfo {title} {Non-hermitian physics},}\
  }\href {\doibase https://doi.org/10.1080/00018732.2021.1876991} {\bibfield
  {journal} {\bibinfo  {journal} {Advances in Physics}\ }\textbf {\bibinfo
  {volume} {69}},\ \bibinfo {pages} {249 -- 435} (\bibinfo {year}
  {2021})}\BibitemShut {NoStop}%
\bibitem [{Eco()}]{EcosystemnoteNote}%
  \BibitemOpen
  \href@noop {} {}\bibinfo {note} {For a large system described by $dX/dt = A
  X$ \cite{May1972Will}, where $X$ is a $N$ column vector and $A = -\mathbb{I}
  + B$ is the matrix at the fixed point, with $I$ being a unity matrix and $B$
  being a random matrix. The variation of the eigenvalues by Eq. 4 in the main
  text yields $\sigma^2 = \langle b_{ij}^2\rangle N^\nu$, thus phase transition
  happens at $\langle b_{ij}^2\rangle N^\nu =1$. Following our theory one may
  consider the stability of coupled large ecosystems.}\BibitemShut {Stop}%
\end{thebibliography}%


\begin{thebibliography}{12}%
\makeatletter
\providecommand \@ifxundefined [1]{%
 \@ifx{#1\undefined}
}%
\providecommand \@ifnum [1]{%
 \ifnum #1\expandafter \@firstoftwo
 \else \expandafter \@secondoftwo
 \fi
}%
\providecommand \@ifx [1]{%
 \ifx #1\expandafter \@firstoftwo
 \else \expandafter \@secondoftwo
 \fi
}%
\providecommand \natexlab [1]{#1}%
\providecommand \enquote  [1]{``#1''}%
\providecommand \bibnamefont  [1]{#1}%
\providecommand \bibfnamefont [1]{#1}%
\providecommand \citenamefont [1]{#1}%
\providecommand \href@noop [0]{\@secondoftwo}%
\providecommand \href [0]{\begingroup \@sanitize@url \@href}%
\providecommand \@href[1]{\@@startlink{#1}\@@href}%
\providecommand \@@href[1]{\endgroup#1\@@endlink}%
\providecommand \@sanitize@url [0]{\catcode `\\12\catcode `\$12\catcode
  `\&12\catcode `\#12\catcode `\^12\catcode `\_12\catcode `\%12\relax}%
\providecommand \@@startlink[1]{}%
\providecommand \@@endlink[0]{}%
\providecommand \url  [0]{\begingroup\@sanitize@url \@url }%
\providecommand \@url [1]{\endgroup\@href {#1}{\urlprefix }}%
\providecommand \urlprefix  [0]{URL }%
\providecommand \Eprint [0]{\href }%
\providecommand \doibase [0]{https://doi.org/}%
\providecommand \selectlanguage [0]{\@gobble}%
\providecommand \bibinfo  [0]{\@secondoftwo}%
\providecommand \bibfield  [0]{\@secondoftwo}%
\providecommand \translation [1]{[#1]}%
\providecommand \BibitemOpen [0]{}%
\providecommand \bibitemStop [0]{}%
\providecommand \bibitemNoStop [0]{.\EOS\space}%
\providecommand \EOS [0]{\spacefactor3000\relax}%
\providecommand \BibitemShut  [1]{\csname bibitem#1\endcsname}%
\let\auto@bib@innerbib\@empty
\bibitem [{\citenamefont {Kravtsov}\ \emph {et~al.}(2015)\citenamefont
  {Kravtsov}, \citenamefont {Khaymovich}, \citenamefont {Cuevas},\ and\
  \citenamefont {Amini}}]{Kravtsov2015Random}%
  \BibitemOpen
  \bibfield  {author} {\bibinfo {author} {\bibfnamefont {V.~E.}\ \bibnamefont
  {Kravtsov}}, \bibinfo {author} {\bibfnamefont {I.~M.}\ \bibnamefont
  {Khaymovich}}, \bibinfo {author} {\bibfnamefont {E.}~\bibnamefont {Cuevas}},\
  and\ \bibinfo {author} {\bibfnamefont {M.}~\bibnamefont {Amini}},\ }\bibfield
   {title} {\bibinfo {title} {A random matrix model with localization and
  ergodic transitions},\ }\href
  {https://doi.org/10.1088/1367-2630/17/12/122002} {\bibfield  {journal}
  {\bibinfo  {journal} {New J. Phys.}\ }\textbf {\bibinfo {volume} {17}},\
  \bibinfo {pages} {122002} (\bibinfo {year} {2015})}\BibitemShut {NoStop}%
\bibitem [{\citenamefont {Tarquini}\ \emph {et~al.}(2016)\citenamefont
  {Tarquini}, \citenamefont {Biroli},\ and\ \citenamefont
  {Tarzia}}]{tarquini2016level}%
  \BibitemOpen
  \bibfield  {author} {\bibinfo {author} {\bibfnamefont {E.}~\bibnamefont
  {Tarquini}}, \bibinfo {author} {\bibfnamefont {G.}~\bibnamefont {Biroli}},\
  and\ \bibinfo {author} {\bibfnamefont {M.}~\bibnamefont {Tarzia}},\
  }\bibfield  {title} {\bibinfo {title} {Level {Statistics} and {Localization}
  {Transitions} of {L}{\textbackslash}'evy {Matrices}},\ }\href
  {https://doi.org/10.1103/PhysRevLett.116.010601} {\bibfield  {journal}
  {\bibinfo  {journal} {Phys. Rev. Lett.}\ }\textbf {\bibinfo {volume} {116}},\
  \bibinfo {pages} {010601} (\bibinfo {year} {2016})}\BibitemShut {NoStop}%
\bibitem [{\citenamefont {Biroli}\ and\ \citenamefont
  {Tarzia}(2021)}]{Biroli2021Levy}%
  \BibitemOpen
  \bibfield  {author} {\bibinfo {author} {\bibfnamefont {G.}~\bibnamefont
  {Biroli}}\ and\ \bibinfo {author} {\bibfnamefont {M.}~\bibnamefont
  {Tarzia}},\ }\bibfield  {title} {\bibinfo {title} {{L\'evy-Rosenzweig-Porter
  random matrix ensemble}},\ }\href
  {https://doi.org/10.1103/PhysRevB.103.104205} {\bibfield  {journal} {\bibinfo
   {journal} {Phys. Rev. B}\ }\textbf {\bibinfo {volume} {103}},\ \bibinfo
  {pages} {104205} (\bibinfo {year} {2021})}\BibitemShut {NoStop}%
\bibitem [{\citenamefont {Aggarwal}\ \emph {et~al.}(2019)\citenamefont
  {Aggarwal}, \citenamefont {Lopatto},\ and\ \citenamefont
  {Yau}}]{aggarwal2019goe}%
  \BibitemOpen
  \bibfield  {author} {\bibinfo {author} {\bibfnamefont {A.}~\bibnamefont
  {Aggarwal}}, \bibinfo {author} {\bibfnamefont {P.}~\bibnamefont {Lopatto}},\
  and\ \bibinfo {author} {\bibfnamefont {H.-T.}\ \bibnamefont {Yau}},\ }\href
  {http://arxiv.org/abs/1806.07363} {\bibinfo {title} {{GOE} {Statistics} for
  {Levy} {Matrices}}} (\bibinfo {year} {2019}),\ \bibinfo {note}
  {arXiv:1806.07363 [math-ph]}\BibitemShut {NoStop}%
\bibitem [{\citenamefont {Mirlin}\ and\ \citenamefont
  {Fyodorov}(1991)}]{mirlin1991universality}%
  \BibitemOpen
  \bibfield  {author} {\bibinfo {author} {\bibfnamefont {A.~D.}\ \bibnamefont
  {Mirlin}}\ and\ \bibinfo {author} {\bibfnamefont {Y.~V.}\ \bibnamefont
  {Fyodorov}},\ }\bibfield  {title} {\bibinfo {title} {Universality of level
  correlation function of sparse random matrices},\ }\href
  {https://doi.org/10.1088/0305-4470/24/10/016} {\bibfield  {journal} {\bibinfo
   {journal} {J. Phys. A: Math. Gen.}\ }\textbf {\bibinfo {volume} {24}},\
  \bibinfo {pages} {2273} (\bibinfo {year} {1991})}\BibitemShut {NoStop}%
\bibitem [{\citenamefont {Fyodorov}\ and\ \citenamefont
  {Mirlin}(1991)}]{fyodorov1991localization}%
  \BibitemOpen
  \bibfield  {author} {\bibinfo {author} {\bibfnamefont {Y.~V.}\ \bibnamefont
  {Fyodorov}}\ and\ \bibinfo {author} {\bibfnamefont {A.~D.}\ \bibnamefont
  {Mirlin}},\ }\bibfield  {title} {\bibinfo {title} {Localization in ensemble
  of sparse random matrices},\ }\href
  {https://doi.org/10.1103/PhysRevLett.67.2049} {\bibfield  {journal} {\bibinfo
   {journal} {Phys. Rev. Lett.}\ }\textbf {\bibinfo {volume} {67}},\ \bibinfo
  {pages} {2049} (\bibinfo {year} {1991})}\BibitemShut {NoStop}%
\bibitem [{\citenamefont {Wei}\ \emph {et~al.}(2019)\citenamefont {Wei},
  \citenamefont {Cheng}, \citenamefont {Xianlong},\ and\ \citenamefont
  {Mondaini}}]{Wei2019Investigating}%
  \BibitemOpen
  \bibfield  {author} {\bibinfo {author} {\bibfnamefont {X.}~\bibnamefont
  {Wei}}, \bibinfo {author} {\bibfnamefont {C.}~\bibnamefont {Cheng}}, \bibinfo
  {author} {\bibfnamefont {G.}~\bibnamefont {Xianlong}},\ and\ \bibinfo
  {author} {\bibfnamefont {R.}~\bibnamefont {Mondaini}},\ }\bibfield  {title}
  {\bibinfo {title} {Investigating many-body mobility edges in isolated quantum
  systems},\ }\href {https://doi.org/10.1103/PhysRevB.99.165137} {\bibfield
  {journal} {\bibinfo  {journal} {Phys. Rev. B}\ }\textbf {\bibinfo {volume}
  {99}},\ \bibinfo {pages} {165137} (\bibinfo {year} {2019})}\BibitemShut
  {NoStop}%
\bibitem [{\citenamefont {Luitz}\ \emph {et~al.}(2015)\citenamefont {Luitz},
  \citenamefont {Laflorencie},\ and\ \citenamefont
  {Alet}}]{Luitz2015Many-body}%
  \BibitemOpen
  \bibfield  {author} {\bibinfo {author} {\bibfnamefont {D.~J.}\ \bibnamefont
  {Luitz}}, \bibinfo {author} {\bibfnamefont {N.}~\bibnamefont {Laflorencie}},\
  and\ \bibinfo {author} {\bibfnamefont {F.}~\bibnamefont {Alet}},\ }\bibfield
  {title} {\bibinfo {title} {Many-body localization edge in the random-field
  {Heisenberg} chain},\ }\href {https://doi.org/10.1103/PhysRevB.91.081103}
  {\bibfield  {journal} {\bibinfo  {journal} {Phys. Rev. B}\ }\textbf {\bibinfo
  {volume} {91}},\ \bibinfo {pages} {081103} (\bibinfo {year}
  {2015})}\BibitemShut {NoStop}%
\bibitem [{\citenamefont {Facoetti}\ \emph {et~al.}(2016)\citenamefont
  {Facoetti}, \citenamefont {Vivo},\ and\ \citenamefont
  {Biroli}}]{Facoetti2016Non-ergodic}%
  \BibitemOpen
  \bibfield  {author} {\bibinfo {author} {\bibfnamefont {D.}~\bibnamefont
  {Facoetti}}, \bibinfo {author} {\bibfnamefont {P.}~\bibnamefont {Vivo}},\
  and\ \bibinfo {author} {\bibfnamefont {G.}~\bibnamefont {Biroli}},\
  }\bibfield  {title} {\bibinfo {title} {From non-ergodic eigenvectors to local
  resolvent statistics and back: {A} random matrix perspective},\ }\href
  {https://doi.org/10.1209/0295-5075/115/47003} {\bibfield  {journal} {\bibinfo
   {journal} {EPL}\ }\textbf {\bibinfo {volume} {115}},\ \bibinfo {pages}
  {47003} (\bibinfo {year} {2016})}\BibitemShut {NoStop}%
\bibitem [{\citenamefont {De~Tomasi}\ and\ \citenamefont
  {Khaymovich}(2022)}]{Tomasi2022Non-Hermitian}%
  \BibitemOpen
  \bibfield  {author} {\bibinfo {author} {\bibfnamefont {G.}~\bibnamefont
  {De~Tomasi}}\ and\ \bibinfo {author} {\bibfnamefont {I.~M.}\ \bibnamefont
  {Khaymovich}},\ }\bibfield  {title} {\bibinfo {title} {{Non-Hermitian
  Rosenzweig-Porter random-matrix ensemble: Obstruction to the fractal
  phase}},\ }\href {https://doi.org/10.1103/PhysRevB.106.094204} {\bibfield
  {journal} {\bibinfo  {journal} {Phys. Rev. B}\ }\textbf {\bibinfo {volume}
  {106}},\ \bibinfo {pages} {094204} (\bibinfo {year} {2022})}\BibitemShut
  {NoStop}%
\bibitem [{\citenamefont {Rosenzweig}\ and\ \citenamefont
  {Porter}(1960)}]{Rosenzweig1960REpulsion}%
  \BibitemOpen
  \bibfield  {author} {\bibinfo {author} {\bibfnamefont {N.}~\bibnamefont
  {Rosenzweig}}\ and\ \bibinfo {author} {\bibfnamefont {C.~E.}\ \bibnamefont
  {Porter}},\ }\bibfield  {title} {\bibinfo {title} {"repulsion of energy
  levels" in complex atomic spectra},\ }\href
  {https://doi.org/10.1103/PhysRev.120.1698} {\bibfield  {journal} {\bibinfo
  {journal} {Phys. Rev.}\ }\textbf {\bibinfo {volume} {120}},\ \bibinfo {pages}
  {1698} (\bibinfo {year} {1960})}\BibitemShut {NoStop}%
\bibitem [{\citenamefont {Giraud}\ \emph {et~al.}(2022)\citenamefont {Giraud},
  \citenamefont {Mac\'e}, \citenamefont {Vernier},\ and\ \citenamefont
  {Alet}}]{Giraud2022Probing}%
  \BibitemOpen
  \bibfield  {author} {\bibinfo {author} {\bibfnamefont {O.}~\bibnamefont
  {Giraud}}, \bibinfo {author} {\bibfnamefont {N.}~\bibnamefont {Mac\'e}},
  \bibinfo {author} {\bibfnamefont {E.}~\bibnamefont {Vernier}},\ and\ \bibinfo
  {author} {\bibfnamefont {F.}~\bibnamefont {Alet}},\ }\bibfield  {title}
  {\bibinfo {title} {Probing symmetries of quantum many-body systems through
  gap ratio statistics},\ }\href {https://doi.org/10.1103/PhysRevX.12.011006}
  {\bibfield  {journal} {\bibinfo  {journal} {Phys. Rev. X}\ }\textbf {\bibinfo
  {volume} {12}},\ \bibinfo {pages} {011006} (\bibinfo {year}
  {2022})}\BibitemShut {NoStop}%
\end{thebibliography}%
\end{document}


\preprint{APS/123-QED}

\title{Supplementary material: Theory of mobility edge and non-ergodic extended phase in coupled random ensembles}
\author{Xiaoshui Lin}
\affiliation{CAS Key Laboratory of Quantum Information, University of Science and Technology of China, Hefei, 230026, China}
\author{Guang-Can Guo}
\affiliation{CAS Key Laboratory of Quantum Information, University of Science and Technology of China, Hefei, 230026, China}
\affiliation{Synergetic Innovation Center of Quantum Information and Quantum Physics, University of Science and Technology of China, Hefei, Anhui 230026, China}
\affiliation{Hefei National Laboratory, University of Science and Technology of China, Hefei 230088, China}
\author{Ming Gong}
\email{gongm@ustc.edu.cn}
\affiliation{CAS Key Laboratory of Quantum Information, University of Science and Technology of China, Hefei, 230026, China}
\affiliation{Synergetic Innovation Center of Quantum Information and Quantum Physics, University of Science and Technology of China, Hefei, Anhui 230026, China}
\affiliation{Hefei National Laboratory, University of Science and Technology of China, Hefei 230088, China}

\date{\today}

\maketitle

\tableofcontents

\section{Characterization of the Rosenzweig-Porter (RP) ensemble}

\begin{figure}[htbp!]
\centering
\includegraphics[width=0.9\textwidth]{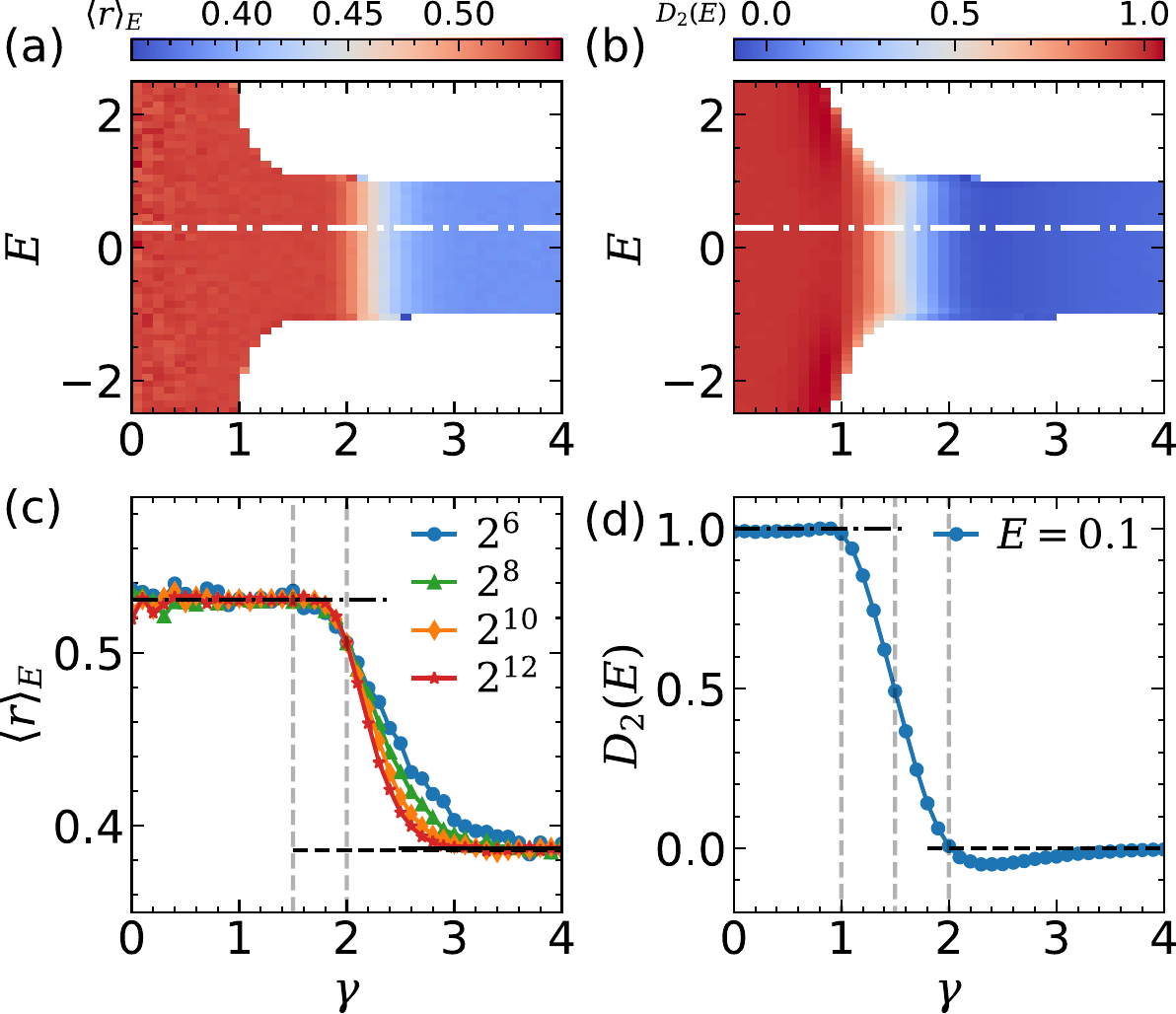}
\caption{Level-spacing ratio $\langle r \rangle_E$ (a) and fractal dimension $ D_2(E)$ (b) against $\gamma$ and $E$ in the RP model. 
The white dashed lines denote $E=0.1$.
(c) A detailed plot of $\langle r\rangle_E$ for various system size $N = 2^6$ - $2^{12}$, exhibiting a critical point at $\gamma = 2$.
(d) The detailed plot of $D_2(E)$ versus $\gamma$, from the extrapolation to $N \rightarrow \infty$ limit.
The distribution function of diagonal terms in Eq. \ref{eq-rp-ensemble} is chosen to be uniform.
}
\label{supfig-rp-uniform}
\end{figure}

\begin{figure}[htbp!]
\centering
\includegraphics[width=0.9\textwidth]{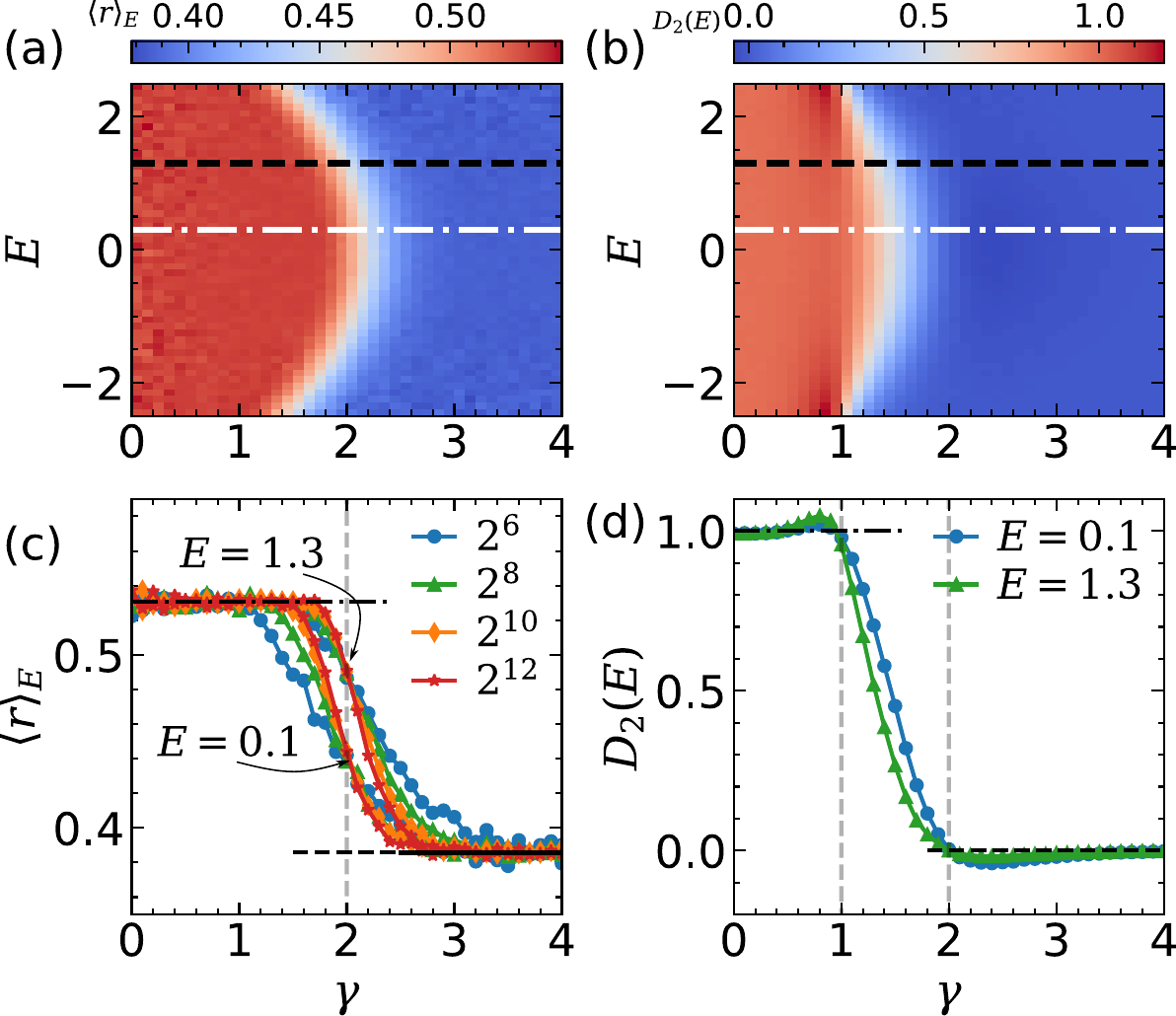}
\caption{The same as that in Fig. \ref{supfig-rp-uniform}. In this plot, the diagonal elements are chosen to Gaussian random variables. 
}
\label{supfig-rp-gaussian}
\end{figure}

We first discuss the Rosenzweig-Porter (RP) ensemble and its related phase transition with a single block. 
The Hamiltonian of this model can be written as \cite{Kravtsov2015Random}
\begin{equation}
    H_{ij} = b_{i}\delta_{ij} + \frac{1}{N^{\gamma/2}}W_{ij}(1 - \delta_{ij}).
    \label{eq-rp-ensemble}
\end{equation}
We assume that the diagonal elements and off-diagonal elements can have different distributions. 
For the diagonal elements, $b_i$ follows $P(b_i) \sim N(0,1)$ for Gaussian distribution or $P(b_i) \sim U(-1, 1)$ for uniform distribution.
For the off-diagonal elements, $W_{ij}$ is a Gaussian random variable with a mean of zero and unit variance. 
It is well-known that there are three phases in the RP ensemble: the ergodic phase (for global coupling), the non-ergodic extended phase (for local coupling), and the localized phase (irrelevant coupling). 
Here, global coupling means that all states are strongly coupled, and local coupling means that only the states with almost degenerate energies are strongly coupled. 
Applying the same treatment developed in the main text, we can obtain the decay rate as 
\begin{equation}
\Gamma(E) = 2\pi\rho(E)N^{1-\gamma},
\end{equation}
based on perturbation theory, which yields $\gamma_\text{ergo} = 1$ for global coupling and $\gamma_\text{AL} = 2$ for local coupling. 
This result is independent of energy, thus, no mobility edge exists. 
To verify this result, we employ the exact-diagonalization method for the model in Eq. \ref{eq-rp-ensemble}, and the results are presented in Fig. \ref{supfig-rp-uniform} (a) - (b) and Fig. \ref{supfig-rp-gaussian} (a) - (b). 
These results indicate that although the numerical results exhibit energy-resolved behaviors at a finite size, this behavior disappears in the large $N$ limit; see Fig. \ref{supfig-rp-uniform} (c) - (d) and Fig. \ref{supfig-rp-gaussian} (c) - (d).
This finite-size behavior, from our expectation, comes from the prefactor $\rho(E)$, which would be irrelevant in the large $N$ limit. Thus, in order to have a mobility edge, one should choose some more complicated off-diagonal couplings, such as the L\'{e}vy distributions \cite{tarquini2016level, Biroli2021Levy, aggarwal2019goe}. 
Nonetheless, the position of the mobility edge can not be tuned. 
Here, we propose an alternative approach to realize tunable mobility edges. 
In our model, the scaling behaviors between energy levels in the overlapped and un-overlapped spectra are totally different, which induces energy-resolved level statistics and the associated mobility edges. 

\clearpage

\section{Phase diagram of the sparse Rosenzwig-Porter ensemble}

\begin{figure}[htbp!]
\centering
\includegraphics[width=0.9\textwidth]{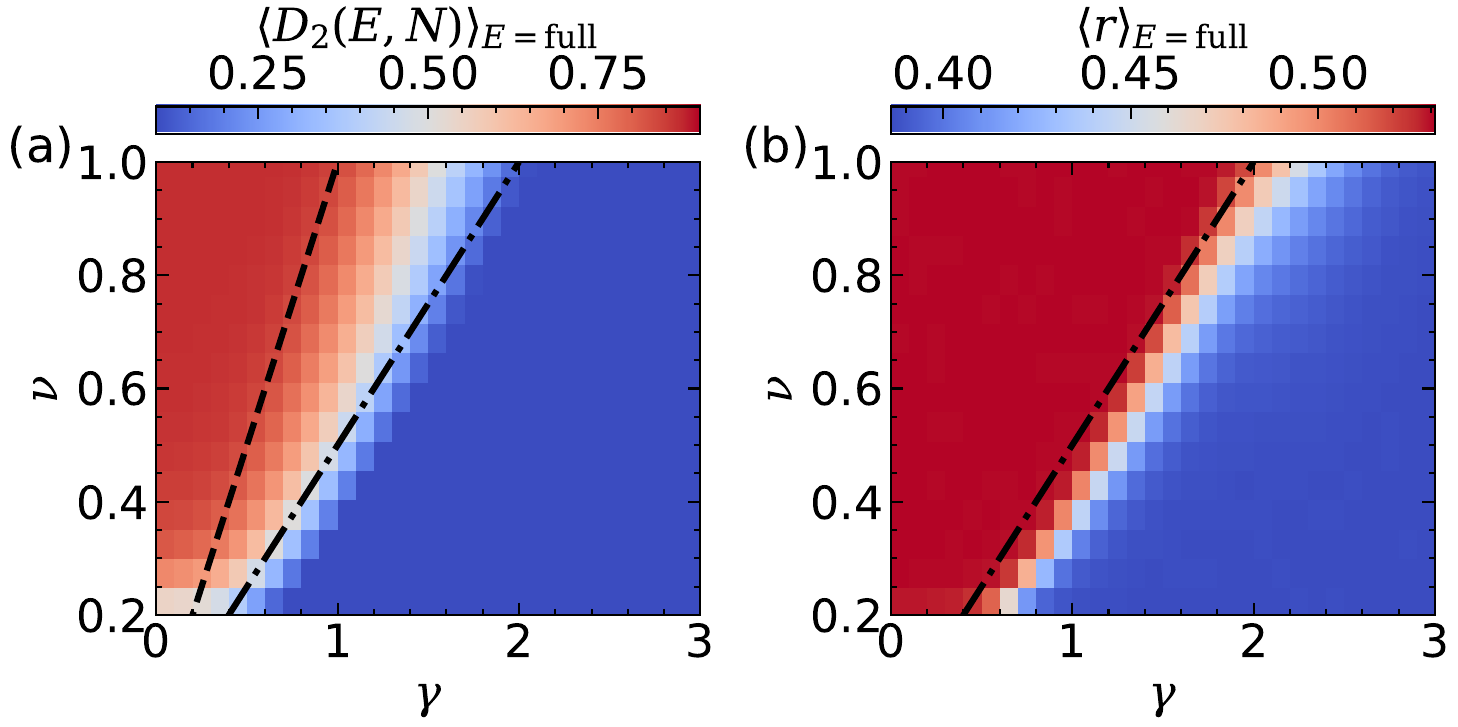}
\caption{(a) Finite-size fractal dimension $D_2(E,N)$ averaged over full spectra for sparse RP ensemble with $N = 2^{11}$. The dashed and dashed-dotted lines represent $\nu = \gamma$ and $\nu = \gamma/2$.
(b) Level-spacing ratio $\langle r\rangle_E$ for sparse RP ensemble. The line denotes $\nu = \gamma/2$.
    }
    \label{supfig-sparse-phase}
\end{figure}

We present the phase diagram of the sparse RP ensemble. 
The Hamiltonian is chosen to be the same as Eq. \ref{eq-rp-ensemble},
but with the distribution of 
\begin{equation}
P(W_{ij}) = (1 - N^{\nu-1})\delta(W_{ij}) + N^{\nu-1} h(W_{ij}).
\end{equation}
Here, $h(x) = \exp(x^2/2)/\sqrt{2\pi}$ .
From the theory in the main text, we expect $\gamma_\text{ergo} = \nu$ and $\gamma_\text{AL} = 2\nu$ for vanished $\beta$ functions. 
The results are presented in Fig. \ref{supfig-sparse-phase}, which agree well with our expectation.
It should be noted that the line with $\gamma = \nu$ has been discussed by Mirlin \text{et. al} \cite{mirlin1991universality,fyodorov1991localization} using super-symmetry method. 
It has been shown that all states are ergodic for $\nu = \gamma \neq 0$.
When it was extrapolated to $\gamma = \nu = 0 $, we will obtain the results studied in many-body localization \cite{Wei2019Investigating,Luitz2015Many-body}.
In this case, the coupling constant $g$ is marginal, and a finite $g_c$ is required for phase transition and energy-resolved mobility edge. 

\section{Derivation of Eq. 4 and Eq. 5 in the main text}

Here, we present a detailed derivation of Eq. 4  and Eq. 5 in the main text, considering the higher-order effect. 
We first consider the exact-solvable limit $g = 0$, the eigenstates and eigenvalue are given by the equation 
\begin{equation}
  H_{\sigma}|i\sigma\rangle = b_{i\sigma}|i\sigma\rangle, 
\end{equation}
with $\sigma\in {0,1}$.
The single-site Green function is given by 
\begin{equation}
    G_{nn,\sigma}^{(0)}(E-i\eta) = 
 \langle n\sigma |\frac{1}{E-\mathcal{H}}|n\sigma \rangle= \frac{1}{E-b_{n\sigma} - i\eta}.
 \label{eq-green-function}
\end{equation}
Then we switch on the coupling $g$, and the Green function can be obtained from the Dyson equation 
\begin{equation}
    G_{nn,\sigma}(E-i\eta) = 
 \frac{1}{E-b_{n\sigma} - i\eta- \Sigma_{n\sigma}(E)} = \frac{E - b_{n\sigma}-R_{n\sigma}}{(E-b_{n\sigma}-R_{n\sigma})^2 + (\eta + \Gamma_{n\sigma})^2} + \frac{i(\eta + \Gamma_{n\sigma})}{(E-b_{n\sigma}-R_{n\sigma})^2 + (\eta + \Gamma_{n\sigma})^2},
 \label{eq-green-function-full}
\end{equation}
where $\Sigma_{n\sigma}(E)$ is the self-energy with $R_{n\sigma} = \Re\Sigma_{n\sigma}(E)$ and $\Gamma_{n\sigma} = \Im\Sigma_{n\sigma}(E)$. 
The imaginary part of the Green function $G_{nn,\sigma}(E-i\eta)$ corresponds to the local density of states $\rho_{nn,\sigma}(E)$, which characterizes the localization properties of the systems.
When the size of the system $N$ approaches infinity, we expect the self-energy to be self-averaged, being independent of the site index $n$.
Thus, we can neglect the index of $n$ and $\rho_{nn,\sigma}(E)$ takes the Breit–Wigner form \cite{Facoetti2016Non-ergodic, Tomasi2022Non-Hermitian}
\begin{equation}
  \rho_{nn,\sigma}(E) = \frac{1}{\pi}\frac{ \Gamma_{\sigma}}{(E-b_{n\sigma}-R_{\sigma})^2 + \Gamma_{\sigma}^2}.
  \label{eq-ldos}
\end{equation}
Here, $\Gamma_{\sigma}$ describes the broadening of the $\rho_{nn,\sigma}(E)$, having the same meaning as that from Fermi's Golden Rule.
$R_\sigma$ is the real part of the self-energy and can be absorbed into the energy shift when $\gamma$ is sufficient large.
Before deriving the expression of $R_\sigma$ and $\Gamma_\sigma$, we first discuss the physical meaning of Eq. \ref{eq-ldos}, which can be recognized as the averaged wave function of states near energy $E = b_{n\sigma} + R_{\sigma}$.
Thus, only the states within energy window $[b_{n\sigma} + R_{\sigma} - \Gamma_{\sigma}/2, b_{n\sigma} + R_{\sigma} + \Gamma_{\sigma}/2]$ are strongly hybridized (by local coupling), which forms the mini-bands.

The expression of $\Gamma_{\sigma}$ is given by the following series
\begin{equation}
    \Gamma(b_{n\sigma}) = 2\pi \sum_{n'\neq n \ \text{or}\ \sigma'\neq \sigma} |\langle n\sigma | \tilde{V} + \tilde{V}(b_{n\sigma}-H_{1-\sigma})^{-1}\tilde{V}^{\dagger} + \tilde{V}(b_{n\sigma}-H_{1-\sigma})^{-1}\tilde{V}^{\dagger}(b_{n\sigma}-H_{\sigma})^{-1}\tilde{V} + \cdots| n'\sigma
    '\rangle|^2\delta(b_{n\sigma} - b_{n'\sigma'}),
\end{equation}
with $\tilde{V} = gV/N^{\gamma/2}$.
The Dirac delta function $\delta(b_{n\sigma} - b_{n'\sigma'})$ inside the summation and the form of $V$ make the interaction mechanism for states in the overlapped and un-overlapped spectra to be different, which induces the mobility edge.
Thus, we first consider $b_{n\sigma}$ in the overlapped spectra, with $\Gamma(b_{n\sigma})$ given by 
\begin{equation}
\begin{aligned}
 \Gamma(b_{n\sigma})/2\pi &= \sum_{n'} |\tilde{V}_{nn'}|^2 \delta(b_{n\sigma} - b_{n',1-\sigma})  + \sum_{n' \neq n} |\sum_m\tilde{V}_{nm}(b_{n\sigma} - b_{m,1-\sigma})^{-1}\tilde{V}_{mn'}^{\dagger} |^2 \delta(b_{n\sigma} - b_{n',\sigma}) + \cdots \\
\end{aligned}
\end{equation}
The odd terms are neglected because $\langle V_{mn}\rangle = 0$.
We replace the summation with the integration and take the ensemble average (i.e. $|\tilde{V}_{nn'}|^2 = \langle \tilde{V}^2\rangle = N^{\nu-1-\gamma}$ and  $|\sum_m\tilde{V}_{nn'}(b_{n\sigma} - b_{m,1-\sigma})^{-1}\tilde{V}_{mn'} |^2 = \sum_m \langle \tilde{V}^2\rangle^2 (b_{n\sigma} - b_{m,1-\sigma})^{-1} $), and we obtain
\begin{equation}
\begin{aligned}
 \Gamma(b_{n\sigma})/2\pi &= N\int dE\rho_{1-\sigma}(E) \langle \tilde{V}^2\rangle \delta(b_{n\sigma} - E)  + N^2\mathcal{P} \int dE dE'\rho_{\sigma}(E) \rho_{1-\sigma}(E') \frac{ \langle \tilde{V}^2\rangle^2} {b_{n\sigma} - E'} \delta(b_{n\sigma} - E) + \cdots \\
 &=N^{1 + \nu-1 - \gamma}\rho_{1-\sigma}(b_{n\sigma}) + N^{2 + 2(\nu-1 - \gamma)}\frac{\rho_{\sigma}(b_{n\sigma})}{U_{1-\sigma}}\ln(\frac{b_{n\sigma}-M_{1-\sigma}+U_{1-\sigma}}{b_{n\sigma}-M_{1-\sigma}-U_{1-\sigma}}) + \cdots.
\end{aligned}
\end{equation}
with $\mathcal{P}$ the Cauchy principal value integration. 
The first term is proportional to $N^{\nu-\gamma}$ and the second term is proportional to $N^{2(\nu-\gamma)}$.  
Thus, we may neglect the higher-order terms when the system is at the large $N$ limit for $\gamma>\nu$, which gives Eq. 4 in the main text.

Then we consider $b_{n\sigma}$ in the un-overlapped spectra. The leading order contribution of the decay rate is given by
\begin{equation}
\begin{aligned}
    \Gamma(b_{n\sigma})/2\pi &= \sum_{n' \neq n} \left[|\sum_m \frac{\tilde{V}_{nm} \tilde{V}_{mn'}^{\dagger}}{(b_{n\sigma} - b_{m,1-\sigma})} |^2  + \cdots\right]\delta(b_{n\sigma} - b_{n',\sigma}) \\
    &= N^2\int dE dE' \rho_{\sigma}(E) \rho_{1-\sigma}(E')\langle \tilde{V}^2\rangle^2\frac{1}{(b_{n\sigma} - E')^2}\delta(b_{n\sigma} - E') + \cdots \\
    &= N^{2 + 2(\nu-1-\gamma)}\rho_{\sigma}(b_{n\sigma})\frac{2U_{1-\sigma}}{(b_{n\sigma}-M_{1-\sigma})^2+ U_{1-\sigma}} + \cdots,
    \end{aligned}
\end{equation}
from which the leading term is proportional to $N^{2(\nu-\gamma)}$ and agrees with Eq. 5 in the main text, and the high-order terms will become more and more irrelevant. The expression of the real part $R_n$ of the self-energy can be obtained with a similar analysis, which can be absorbed into the energy shift when $\gamma >\nu$.

\clearpage

\section{Finite-size analysis of the phase boundaries}

\begin{figure}[htbp!]
\centering
\includegraphics[width=0.9\textwidth]{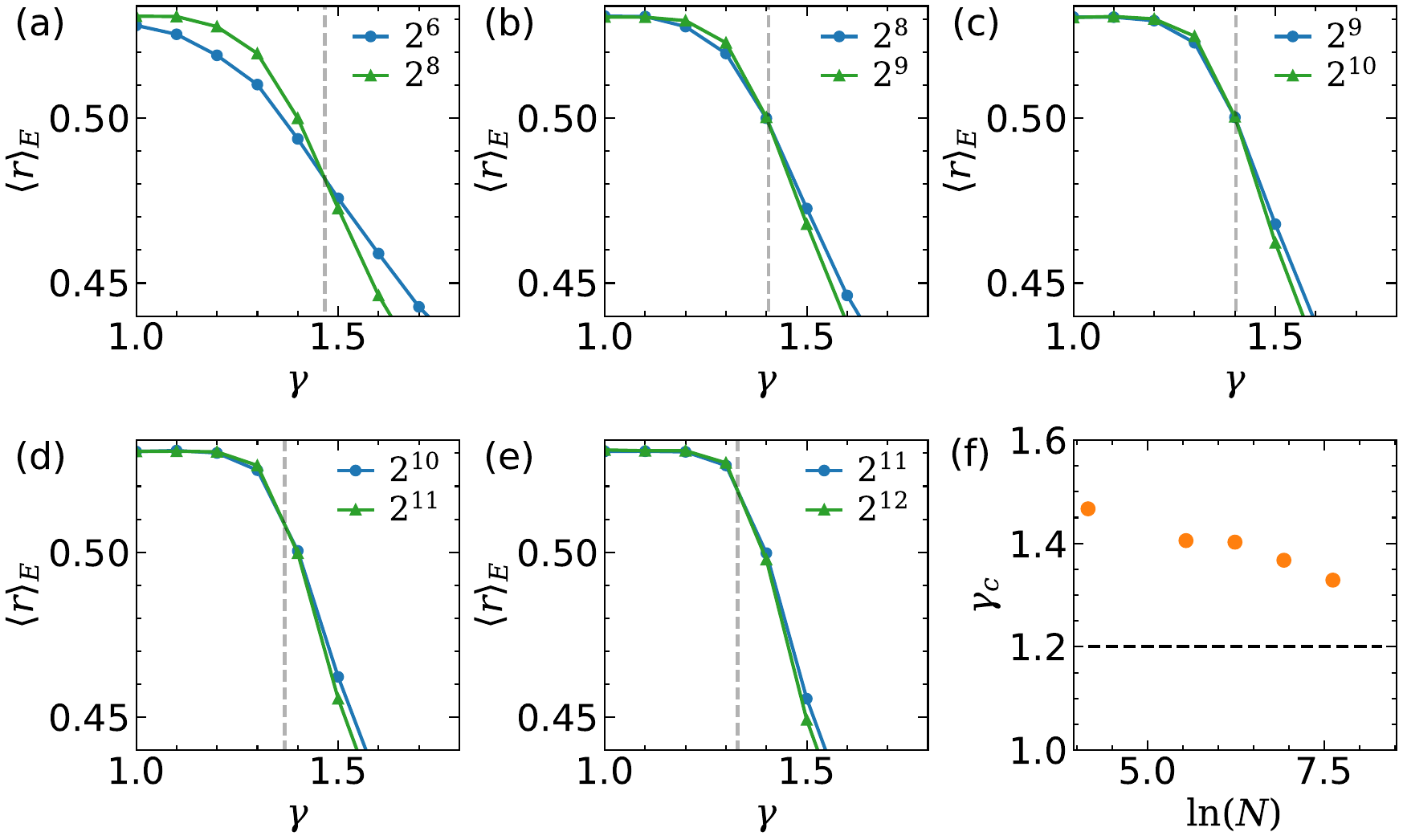}
\caption{ (a)-(e) Level-spacing ratio averaged over energy window $E\in [0,1] $ (in the overlapped spectra) and $\nu = 0.6$ for different system sizes. The dashed grey lines are the apparent cross points $\gamma_c$.
(f) $\gamma_c$ vs. $\ln(N)$. The black dashed line is the theoretical expectation. In this case, $\gamma_c = 2\nu = 1.2$ from theoretical prediction. 
    }
    \label{supfig-scale-cross-point-overlap}
\end{figure}

\begin{figure}[htbp!]
\centering
\includegraphics[width=0.9\textwidth]{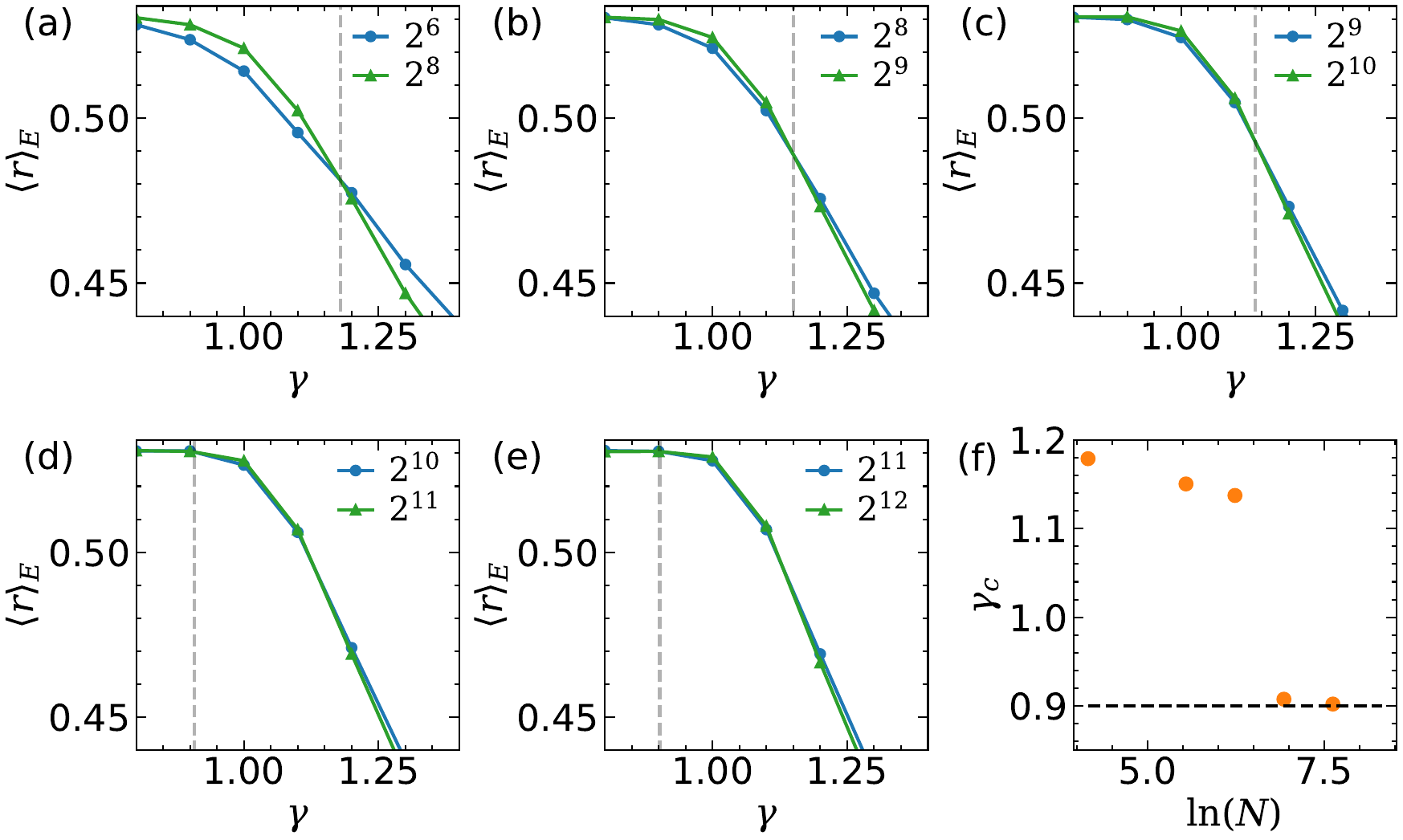}
\caption{ The same with Fig. \ref{supfig-scale-cross-point-overlap} 
for results in the un-overlapped spectra. In this case, $\gamma_c = 3\nu/2 = 0.9$ from theoretical prediction. }
    \label{supfig-scale-cross-point-unoverlap}
\end{figure}

In the main text, we have presented the results of the level-spacing ratio for states with $\nu = 0.6$, finding that the phase boundary is slightly different from the theoretical prediction. This comes from the finite size effect. In Fig. \ref{supfig-scale-cross-point-overlap} and Fig. \ref{supfig-scale-cross-point-unoverlap}, we analysis the cross point for various sizes, showing that with the increasing of system size $N$, the cross point and the associated $\gamma_c$ will approaching the theoretical value when $N\rightarrow \infty$; see  Fig. \ref{supfig-scale-cross-point-overlap} (f) and Fig. \ref{supfig-scale-cross-point-unoverlap} (f). 

\clearpage

\section{Coupled random matrices with Gaussian unitary ensemble (GUE)}

\begin{figure}[htbp!]
\centering
\includegraphics[width=0.9\textwidth]{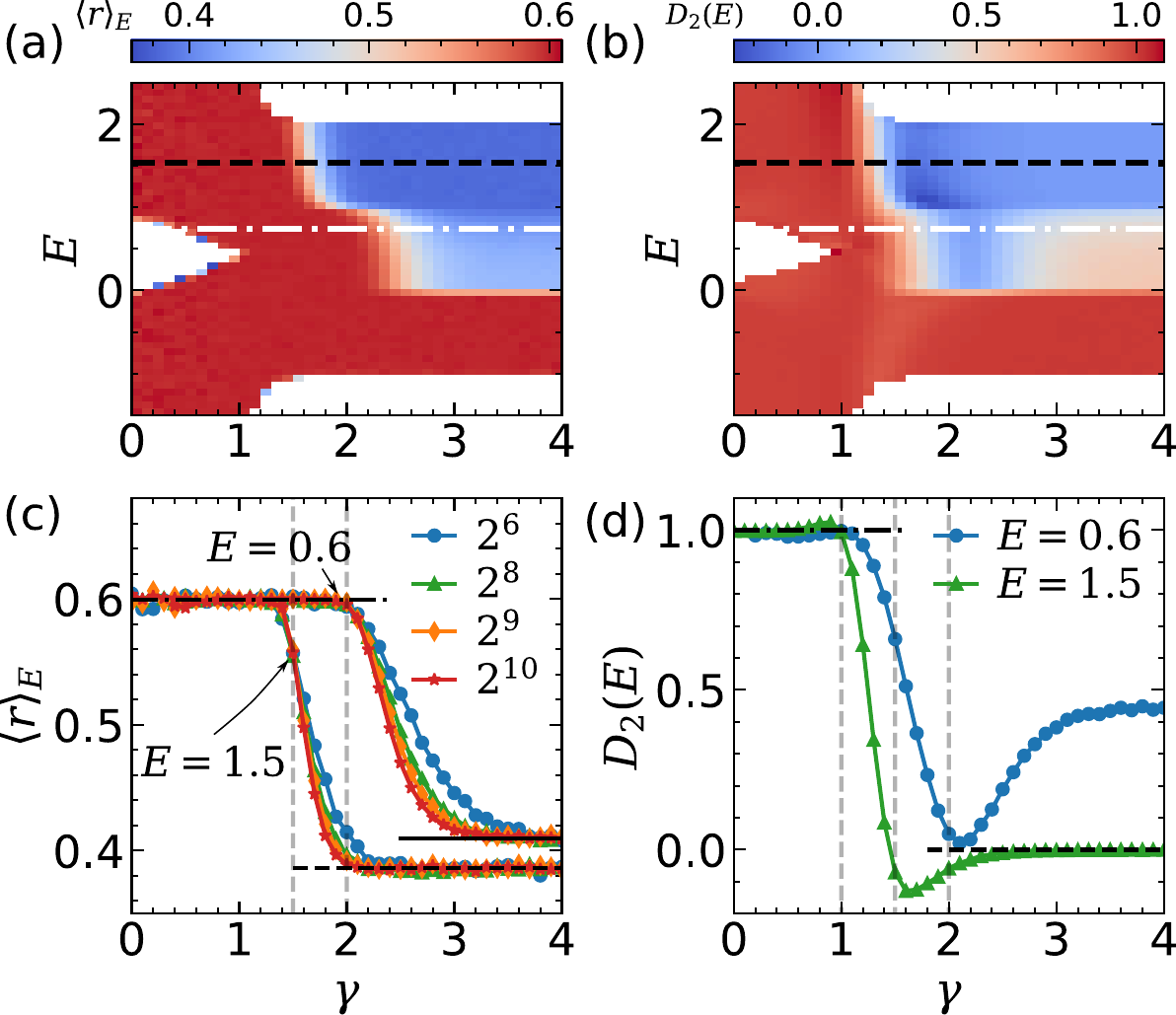}
\caption{Energy-resolved level-spacing ratio $\langle r \rangle_E$ (a) and fractal dimension $D_2(E)$ (b) against $\gamma$ and $E$ for coupled random matrices with $H_0$ being PE and $H_1$  being GUE. The dashed lines denote $E=0.6$ and 1.5, respectively. 
(c) A detailed plot of $\langle r\rangle_E$ with critical point at $\gamma = 2$, which is independent of system sizes. 
(d) The detailed plot of $D_2(E)$ versus $\gamma$.
The elements of $H_1$ are chosen to have zero mean and variance of $1/\sqrt{2N}$, in order to make the spectra overlap in the interval $[0,1]$. 
}
\label{supfig-couple-pe-gue}
\end{figure}

\begin{figure}[htbp!]
\centering
\includegraphics[width=0.9\textwidth]{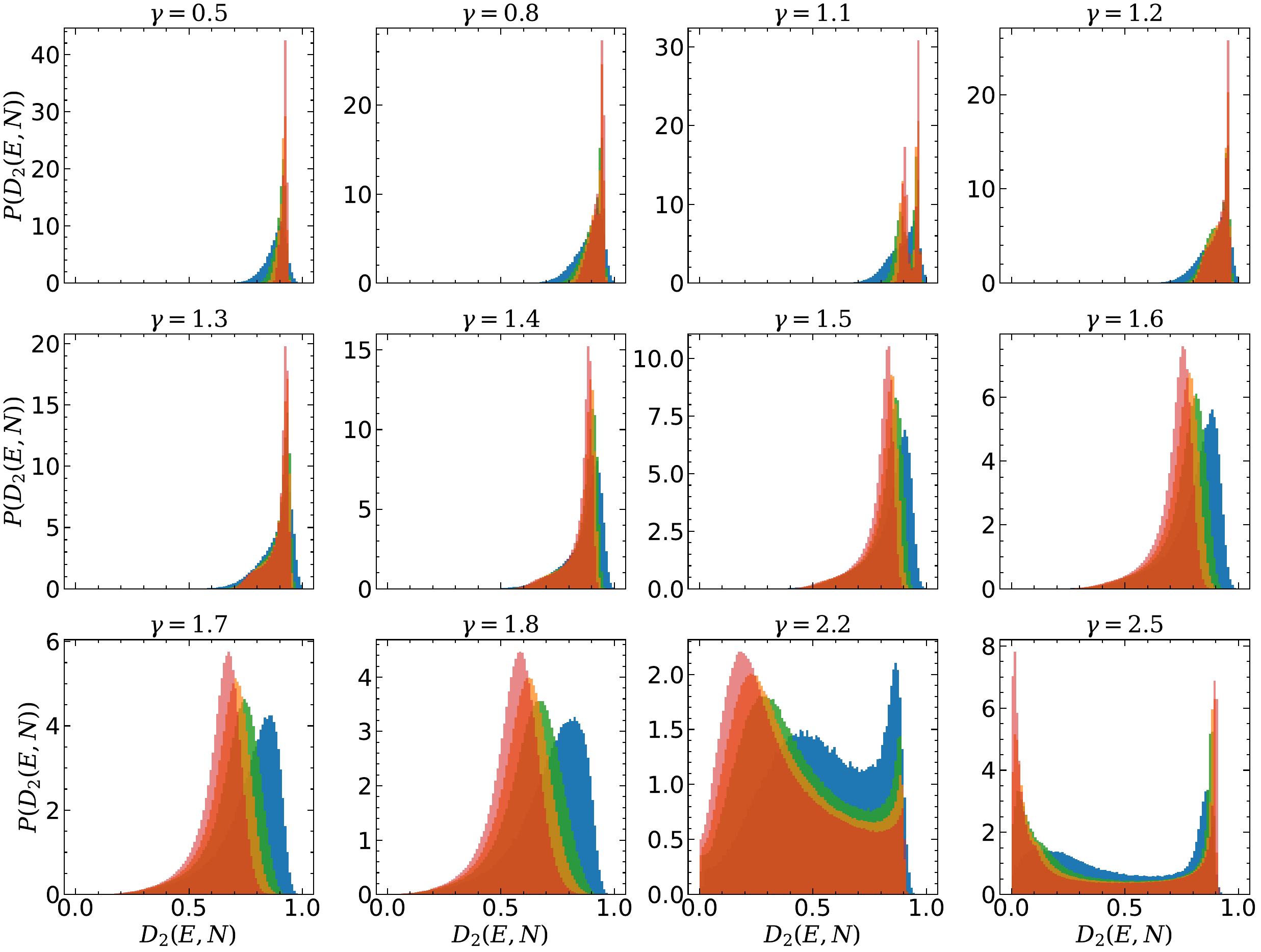}
\caption{ 
The distribution of $D_2(E, N)$ at different $\gamma$ and system sizes $N$ for states with $|E-0.5|<0.5$ in the overlapped spectra.
We use $H_0$ as PE and $H_1$ as GUE.
We use $N=2^6$ (blue), $N=2^8$ (green), $N=2^9$ (yellow), and $N=2^{10}$ (red) in all panels.
The distribution function of the coexistence phase exhibits two peaks in the irrelevant regime and one peak in the relevant regimes. 
}
    \label{supfig-distrib-pe-gue}
\end{figure}

In the main text, we have discussed coupled random matrices with $H_0$ and $H_1$ being PE for mobility edge and NNE phase.
These phenomena also exist when one of the random matrices in the diagonal block is replaced by GOE.
Here, we use $H_0$ belonging to PE, $H_1$ belonging to GUE, and the elements of $V$ being complex numbers with Gaussian distribution and $\nu = 1$. 
From our theory, we expect $\gamma_\text{ergo} = 1$ for overlapped and un-overlapped spectra and $\gamma_\text{AL} = 3/2$ ($\gamma_\text{AL} = 2$) for un-overlapped (overlapped) spectra. 
We presented the results in Fig. \ref{supfig-couple-pe-gue}. 
The critical point of level-spacing ratio $\langle r \rangle_E $ happens at $\gamma = 3/2$ for un-overlapped spectra, and at $\gamma = 2$ for overlapped spectra.
We emphasize that $\langle r \rangle_E \neq 0.386$ for overlapped spectra at $\gamma>2$ is due to the coexistence of spectra from PE and GOE in the irrelevant regime, following the value given in Ref. \citep{Rosenzweig1960REpulsion, Giraud2022Probing}. 
This coexistence can also be viewed from the fractal dimension $D_2(E,N)$. We find $D_2(E) \sim 0.45 $ for overlapped spectra at $\gamma>2$. This value of $ D_2(E) $ does not indicate an NEE phase, which can be reflected from the distribution of $ D_2(E, N)$ in Fig. \ref{supfig-distrib-pe-gue}.
The distribution function of $D_2(E,N)$ exhibits two peaks when $\gamma>2$ with one peak approaching unity and another approaching zero.
When $1<\gamma<2$, we find the distribution function only exhibits one peak for the overlapped spectra, indicating that all states are fractal.
Therefore, we demonstrate that by coupling between PE and GUE, we can also realize the NEE phase and mobility edge in the intermediate regime with a proper value of $\gamma$. 

\clearpage

\bibliography{ref}